\documentclass[aps,prd,twocolumn,superscriptaddress,nofootinbib,floatfix]{revtex4-2}
\usepackage{graphicx}
\usepackage{mathrsfs}
\usepackage{bm}
\usepackage{amssymb}
\usepackage{amsmath}
\usepackage{epsfig}

\usepackage{color}
\usepackage{mathtools}
\usepackage{cases}
\usepackage{booktabs}
\usepackage{natbib}

\usepackage{tabularx}
\usepackage{placeins}

\usepackage[
  colorlinks=true,
  linkcolor=blue,
  citecolor=blue,
  urlcolor=red,
  filecolor=blue,
  anchorcolor=blue,
  menucolor=blue,
  breaklinks=true,
]{hyperref}
\usepackage{orcidlink}

\begin{document}

\title{Rapid Multiparameter Gravitational-Wave Tests of General Relativity with Neural Posterior Estimation}

\author{Hong Su\orcidlink{0009-0000-3277-2986}}
% \email
% {suhong@whu.edu.cn}
\affiliation{School of Physics, Wuhan University, Wuhan 430072, China}

\author{Tao Yang\orcidlink{0000-0002-2161-0495}}
\email[Corresponding author:~]
{yangtao@whu.edu.cn}
\affiliation{School of Physics, Wuhan University, Wuhan 430072, China}

\author{He Wang\orcidlink{0000-0002-1353-391X}}
\email[Corresponding author:~]
{hewang@ucas.ac.cn}
\affiliation{Taiji Laboratory for Gravitational Wave Universe (Beijing/Hangzhou), University of Chinese Academy of Sciences (UCAS), Beijing 100049, China}
\affiliation{International Centre for Theoretical Physics Asia-Pacific (ICTP-AP), University of Chinese Academy of Sciences (UCAS), Beijing 100049, China}

\date{\today}

\begin{abstract}
Multiparameter gravitational-wave tests of general relativity (GR) are limited by strong correlations among post-Newtonian (PN) phase deformations and the computational cost of Bayesian inference. 
We present DINGO-TGR, a neural posterior estimation framework that jointly infers binary-source parameters and six PN phase deformations within the TIGER parametrization, using IMRPhenomXPHM waveforms with spin precession and higher-order modes. 
Importance sampling reweights the neural proposal samples, while principal component analysis (PCA) identifies two well-constrained combinations of PN deformations. We validate the framework using 16 zero-noise GR injections and 12 beyond-GR injections, recovering GR-consistent posteriors in the former and leading-mode shifts consistent with the injected deformations in the latter. 
For GW150914 and GW250114, our results agree with independent \texttt{bilby-tgr} analyses using \texttt{dynesty}. After training, inference including importance sampling takes approximately 38 minutes and 4.4 hours, respectively, compared with 177 and 238 hours for conventional sampling. 
GW150914 exhibits a mild offset from GR in the leading PCA mode, whereas GW250114 is consistent with GR. 
To assess the significance of the GW150914 offset, we analyze 1000 GW150914-like GR injections with independent Gaussian-noise realizations and fixed detector power spectral densities. 
The resulting empirical upper-tail $p$-value is $11.5\%$, indicating that the observed offset does not reject the adopted GR null model at the 5\% significance level. Our results demonstrate the potential of importance-sampling-corrected neural inference for efficient multiparameter GR tests and large-scale injection-based significance assessments.

\end{abstract}

\maketitle
%\vspace*{0. 5cm}
%\maketitle
%\baselineskip 7. 8mm 

%\tableofcontents
\allowdisplaybreaks

\section{Introduction}
The first direct detection of gravitational waves (GW) from a binary black hole (BBH) merger in 2015 opened a new avenue for testing general relativity (GR)~\cite{LIGOScientific:2016aoc,LIGOScientific:2016vlm}.
BBH mergers provide natural laboratories for probing nonlinear gravitational dynamics in the strong-field regime and searching for signatures of beyond-GR (BGR) physics~\cite{LIGOScientific:2016lio}.

Parametrized tests of compact-binary inspirals probe departures from GR through deviations in the GW phase evolution.
The adiabatic inspiral phase can be well described by post-Newtonian (PN) theory~\cite{Blanchet:2006jqj}.
In PN theory, the amplitude and phase of GW signals are expanded in powers of the characteristic orbital velocity $v$ of the binary~\cite{Blanchet:1995fg}.
The sensitivity of inspiral tests to phase evolution motivates introducing parametrized deviations into the PN phase coefficients~\cite{Tahura:2019dgr,Mishra:2010tp,Agathos:2013upa,Mehta:2022pcn}.
The inspiral phase has been calculated through 4.5PN order for quasicircular, nonspinning compact binaries~\cite{Blanchet:2023bwj}.
Here, we restrict the parametrized deformations to terms through 3.5PN order, as commonly adopted in tests of GR.

PN phase coefficients encode nonlinear effects in compact-binary dynamics at different PN orders. Because these effects can differ between GR and alternative theories, phenomenological deformations of the coefficients are sensitive probes of departures from GR.
This parametrization is termed ``null'' because all deviations vanish identically in GR.
Thus, any statistically significant deviation from the null hypothesis may indicate physics beyond GR.
We adopt the fractional PN phase deformations of the Test Infrastructure for General Relativity (TIGER) framework~\cite{Agathos:2013upa,Meidam:2017dgf} and use the frequency-domain waveform approximant IMRPhenomXPHM~\cite{Pratten:2020ceb}.

Efficient posterior sampling is essential for GW parameter inference.
Bayesian-inference software such as \texttt{Bilby}~\cite{Ashton:2018jfp}, coupled with nested samplers such as \texttt{dynesty}~\cite{Speagle:2019ivv}, provides a conventional approach to this task.
However, repeated sampling of our 21-parameter model, which comprises 15 standard GW parameters and six PN deformation parameters, is computationally demanding, particularly for large injection campaigns.
This challenge is especially pronounced for high signal-to-noise-ratio (SNR) events observable with next-generation ground-based detectors and for long-duration signals from binary-neutron-star coalescences.
The cumulative cost of Markov chain Monte Carlo (MCMC) and nested sampling can therefore limit large-scale statistical studies.

Deep learning has developed rapidly within the GW community and provides an effective way to reduce this computational cost.
We focus on its application to GW parameter estimation.
Recent advances in simulation-based inference (SBI)~\cite{Cranmer:2019eaq,Liang:2025yjw}, particularly neural posterior estimation (NPE)~\cite{Chua:2019wwt}, have substantially reduced the time required for GW parameter estimation. The Deep Inference for Gravitational-wave Observations (DINGO) framework exemplifies these advances~\cite{Green:2020hst,Dax:2021tsq,Dax:2022pxd,Kofler:2025dux}.
Neural methods have also been applied to tests of GR~\cite{Xie:2024ubm,Xie:2025voe,Zhang:2026okw} and searches for signals with PN phase deviations~\cite{Wang:2024oei}.
Related applications include ringdown inference and Bayesian model comparison~\cite{Crisostomi:2023tle,Srinivasan:2024uax,Pacilio:2024qcq}.
These methods also support analyses of lensed GW events~\cite{Chan:2025pdf,Chan:2025kyu} and constraints on cosmological parameters, particularly the Hubble parameter~\cite{Stachurski:2023ntw,Leyde:2023iof}.
Neural estimators have been developed for source-parameter inference in glitch-contaminated data~\cite{Xiong:2024gpx}.
Amortized inference may also benefit long-duration signals, such as extreme-mass-ratio inspirals.

DINGO uses normalizing flows~\cite{Kobyzev:2019ydm,Papamakarios:2019fms}, an architecture also adopted by \texttt{BilbyFlow}~\cite{Pinchbeck:2026hup}. These models are trained on simulated GW data to learn the mapping from observed signals to the source-parameter posterior. Training minimizes the negative log-likelihood of the training parameters conditioned on the data. Once trained, the network infers GW source parameters for an input signal. Previous DINGO studies report accuracy comparable to that of conventional sampling methods, with neural posterior samples generated in seconds to minutes per event~\cite{Green:2020hst,Dax:2021tsq}.
Because the neural proposal only approximates the target posterior, DINGO uses importance sampling (IS) to reweight samples drawn from the proposal using the target likelihood and prior. The resulting weights correct for differences between the proposal and the target Bayesian posterior; this method is denoted DINGO-IS~\cite{Dax:2022pxd}. With adequate proposal coverage and effective sample size, this procedure enables efficient estimation of target-posterior quantities and is well suited to large-scale GW inference.

We extend DINGO's waveform-generation interface~\cite{dingo_gw_repo} to pass PN deformation parameters to the existing TIGER implementation in LALSuite.
Our workflow, hereafter DINGO-TGR, jointly infers binary parameters and six PN deformations using NPE followed by importance sampling.

Parametrized inspiral tests of GR can be implemented in several ways.
The simplest approach is the single-parameter test, in which one deformation parameter is varied at a time while all the remaining testing parameters are fixed at zero.
This strategy has been widely adopted by the LIGO--Virgo--KAGRA (LVK) Collaboration in analyses of observed GW signals~\cite{LIGOScientific:2016lio,LIGOScientific:2018dkp,LIGOScientific:2019fpa,LIGOScientific:2021sio}.
However, departures from GR can affect several PN phase coefficients simultaneously, potentially producing correlated deviations that single-parameter tests do not fully characterize~\cite{Mahapatra:2025cwk}.
Multiparameter tests address this limitation by varying several deformation parameters jointly while fixing the remaining ones to zero.
Lower-order fractional PN deformation parameters are degenerate with intrinsic binary parameters such as chirp mass and symmetric mass ratio~\cite{Mahapatra:2025cwk}. Given these degeneracies and the sensitivity of current GW detectors, we vary only six higher-order PN deviation parameters from 1.5PN to 3.5PN.
Strong parameter degeneracies can nevertheless leave the marginal posteriors of individual deformation parameters weakly constrained, as illustrated by the multiparameter analysis of GW150914~\cite{LIGOScientific:2016lio}.
Related classification results show that different phase-deformation families can remain difficult to distinguish even with intrinsic source parameters fixed to their injected values~\cite{Heisenberg:2026Identifiability}.

Weak constraints on individual deformation parameters do not preclude informative constraints on their combinations.
Early Fisher-matrix studies identified well-measured linear combinations of PN coefficients~\cite{Pai:2012mv}.
A complementary waveform-space approach uses singular value decomposition (SVD) to construct deformation directions orthogonal to GR parameter variations~\cite{Seymour:2026bjg}.
Posterior-based principal component analysis (PCA) has also been applied to LISA simulations~\cite{Niu:2023kzz}, LVK observations~\cite{Saleem:2021nsb,LIGOScientific:2026fcf}, and GW propagation tests~\cite{Ma:2024kkz}.
For GW170817, posterior covariance diagonalization constrained combinations of PN deviations~\cite{Shoom:2021mdj}.
Following Ref.~\cite{Mahapatra:2025cwk}, we apply PCA to the joint deformation-parameter posterior to identify the best-constrained linear combinations of the deformation parameters. Focusing on a small set of well-constrained modes, such as the leading and subleading modes, reduces the effective dimensionality of the BGR parameter space. The constraints on these combinations encoded in the original joint posterior are preserved. We combine this posterior compression with rapid NPE to perform null tests of GR, validate the approach through extensive injection studies, and apply it to real GW observations. This combination enables rapid injection studies and statistical assessment of apparent deviations from GR.

This paper is organized as follows.
Section~\ref{sec:method} presents our framework for joint inference and PCA analysis of PN deformations.
Section~\ref{sec:results} assesses posterior recovery using zero-noise GR and BGR injections and applies the framework to GW150914 and GW250114.
It also uses 1000 GW150914-like GR injections with independent Gaussian-noise realizations to assess how unusual the GW150914 offset is under the adopted noise model.
Our conclusions are given in Sec.~\ref{sec:conclusion}.

\section{Method}
\label{sec:method}
In GR, the early inspiral of compact binary coalescence can be well approximated by the PN formalism~\cite{Blanchet:2006jqj}, in which the GW amplitude and phase are expanded as power series in the characteristic orbital velocity $v$ of the system.
For a quasicircular compact binary, the frequency-domain inspiral phase of the dominant $(2,2)$ harmonic, computed using the stationary-phase approximation~\cite{Sathyaprakash:1991mt,Damour:2000gg}, takes the following form in GR:
\begin{equation}
\begin{split}
\Phi_{\mathrm{insp}}(f)
&=
2\pi f t_c-\phi_c-\frac{\pi}{4}\\
&\quad+\frac{3}{128\eta v^5}
\left[
\sum_{k=0}^{7}
(\phi_k+\phi_{k\ell}\ln v)v^k
+\mathcal{O}(v^8)
\right],
\end{split}
\label{eq:GR_inspiral_phase}
\end{equation}
where $t_c$ and $\phi_c$ are the time and phase of coalescence, and $\eta$ is the symmetric mass ratio, defined as the reduced mass divided by the total mass.
The prior bounds on $t_c$ are relative to the trigger time.
The characteristic orbital velocity can be expressed as $v = \left(\pi M f\right)^{1/3}$, where $f$ is the GW frequency and $M$ is the detector-frame (redshifted) total mass of the binary.
The index $k$ denotes the $k/2$-PN-order correction to the GW phase.

The PN coefficients, denoted by $\phi_k$ (nonlogarithmic) and $\phi_{k\ell}$ (logarithmic), are functions of the component masses and spins of a compact binary system; they encode nonlinear effects in GR.
We adopt geometrized units ($G=c=1$) unless otherwise stated.

\subsection{Multiparameter tests of GR with TIGER}
We adopt TIGER~\cite{Agathos:2013upa}, a data-analysis pipeline for parametrized multiparameter tests of GR with GW from coalescing compact binaries.
TIGER parametrizes possible departures from GR by introducing phenomenological, dimensionless fractional deviation parameters $\delta\hat{\phi}_b$ that deform the PN phase coefficients of the waveform:
\begin{equation}
\phi_b \longrightarrow \phi_b^{\mathrm{GR}}\,(1+\delta\hat{\phi}_b),
\label{eq:deformation}
\end{equation}
where the subscript $b$ collectively denotes both the nonlogarithmic and logarithmic PN coefficients introduced above.
In Eq.~\eqref{eq:deformation}, the coefficients $\phi_b^{\mathrm{GR}}$ denote the nonspinning parts~\cite{Roy:2025gzv}.
This parametrization is theory-agnostic: all $\delta\hat{\phi}_b$ vanish identically in GR, whereas in alternative theories of gravity one or more of them can take nonzero values.
A direct measurement of the $\delta\hat{\phi}_b$ with GW data therefore constitutes a null test of GR, with GR as the null hypothesis~\cite{Mahapatra:2025cwk}.

For the nonzero PN coefficients in the GR inspiral phase through 3.5PN order, the complete set of fractional deformation parameters is
\begin{align}
\delta\vec{\phi}_k &= \{\delta\hat{\phi}_0, \delta\hat{\phi}_2, \delta\hat{\phi}_3, \delta\hat{\phi}_4, \delta\hat{\phi}_6, \delta\hat{\phi}_7\}, \\
\delta\vec{\phi}_{k\ell} &= \{\delta\hat{\phi}_{5\ell}, \delta\hat{\phi}_{6\ell}\}.
\end{align}
We do not introduce a fractional deviation for the nonlogarithmic term at 2.5PN order, since $\delta\hat{\phi}_5$ can be absorbed into the coalescence phase $\phi_c$.

We vary only the six deviation parameters from 1.5PN through 3.5PN,
\begin{equation}
\vec{\theta}_{\mathrm{D}} = \{\delta\hat{\phi}_3, \delta\hat{\phi}_4, \delta\hat{\phi}_{5\ell}, \delta\hat{\phi}_6, \delta\hat{\phi}_{6\ell}, \delta\hat{\phi}_7\},
\label{eq:theta_D}
\end{equation}
while fixing all lower-order deviations $\{\delta\hat{\phi}_{-2},\delta\hat{\phi}_0,\delta\hat{\phi}_1,\delta\hat{\phi}_2\}$ to zero.
The low-order deviations are strongly degenerate with intrinsic binary parameters such as the chirp mass (0PN) and the symmetric mass ratio (1PN)~\cite{Mahapatra:2025cwk,Sanger:2024axs}. 

For both signal generation and parameter estimation, we employ the waveform model IMRPhenomXPHM~\cite{Pratten:2020ceb, Roy:2025gzv}, which includes spin precession and higher-order multipoles. 
We allow generic spin orientations, parametrized by spin magnitudes $a_1,a_2$, tilt angles $\theta_1,\theta_2$ relative to the orbital angular momentum, and azimuthal angles $\phi_{12},\phi_{JL}$.
The PN deviations of Eq.~\eqref{eq:deformation} are applied to the inspiral phase and propagated to the retained higher-order modes through the approximate harmonic phase mapping~\cite{Roy:2025gzv},
\begin{equation}
\Phi_{\ell m}^{\mathrm{insp}}(f)
\simeq \frac{m}{2}\,
\Phi_{22}^{\mathrm{insp}}\!\left(\frac{2f}{m}\right),
\qquad m>0.
\label{eq:higher_mode_phase_mapping}
\end{equation}
The corresponding mode-dependent inspiral phase corrections are given in Eq.~(12) of Ref.~\cite{Roy:2025gzv}.
The merger--ringdown sector is kept as in GR.

\subsection{Neural posterior estimation and importance sampling}
\label{subsec:npe}
\begin{figure*}[!htbp]
	\centering
	\includegraphics[width=1.0\textwidth]{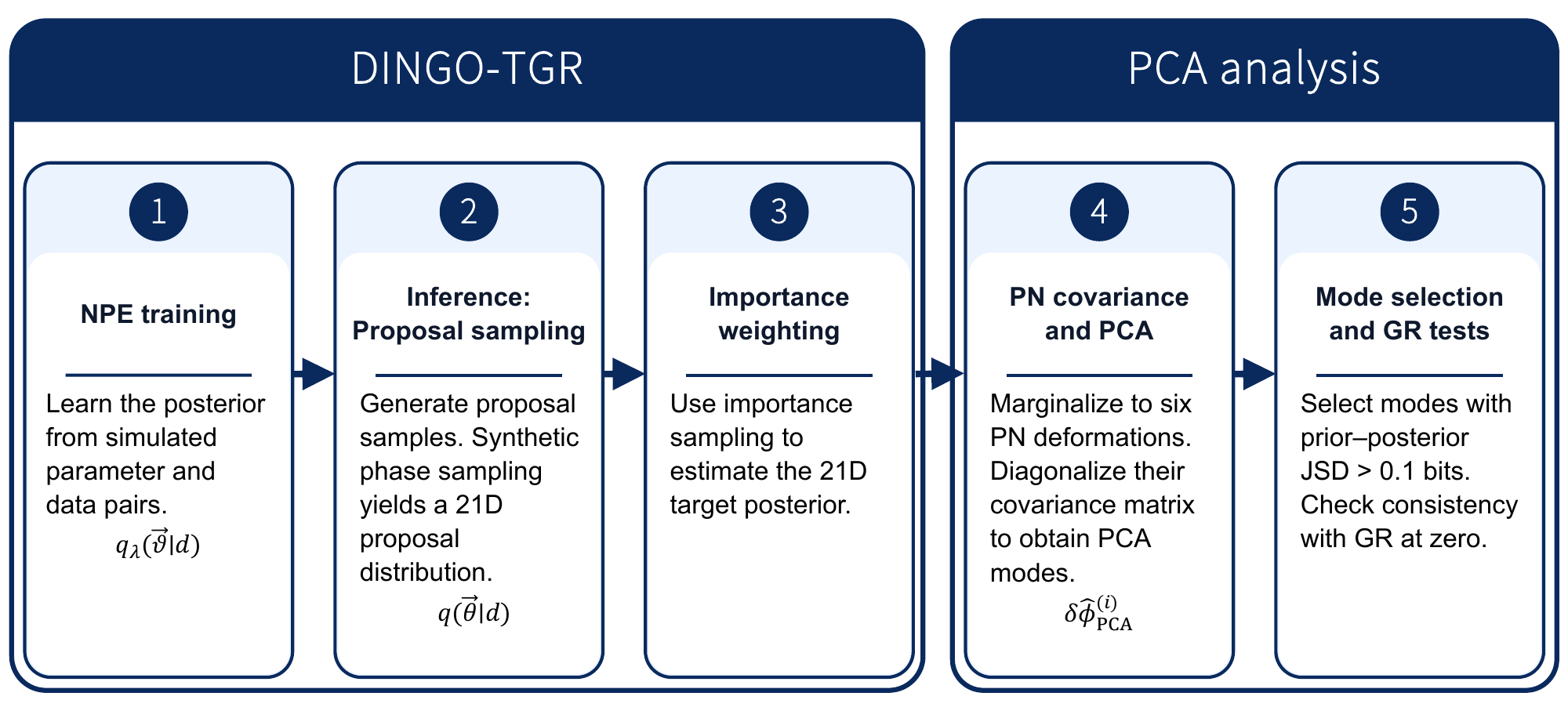}
	\caption{Overview of the inference and PCA workflow. A neural posterior estimator $q_\lambda(\vec{\vartheta}\mid d)$ is trained on simulated data and generates 20-dimensional proposal samples. Synthetic phase sampling extends these to 21 dimensions, after which importance sampling reweights them to approximate the target posterior. After marginalizing over the standard GW parameters, we diagonalize the importance-weighted covariance matrix of the PN deformations to construct the PCA modes. Finally, modes whose prior and posterior distributions have a Jensen--Shannon divergence greater than 0.1 bits are retained for the GR tests.}
	\label{fig:method_workflow}
\end{figure*}

Figure~\ref{fig:method_workflow} summarizes the full workflow. Steps 1--3 are described here, while steps 4--5 are detailed in the next subsection.

Jointly sampling the deviation parameters of Eq.~\eqref{eq:theta_D} with the standard binary source parameters increases the computational cost of MCMC and nested sampling, particularly when repeated over large injection campaigns.
We therefore adopt SBI and perform NPE with the DINGO framework~\cite{Green:2020dnx,Dax:2021tsq}, as implemented in the \texttt{DINGO-gw} software.
A network trained on simulated data rapidly generates proposal samples for new events within its training domain, reducing the need for repeated conventional sampling.

DINGO employs normalizing flows to learn an invertible, data-dependent map from a simple base distribution to a conditional posterior approximation.
Denoting the full 21-dimensional parameter set by
\begin{equation}
\vec{\theta} = \{\vec{\theta}_{\mathrm{GR}},\, \vec{\theta}_{\mathrm{D}}\},
\label{eq:full_parameter_set}
\end{equation}
with $\vec{\theta}_{\mathrm{D}}$ given in Eq.~\eqref{eq:theta_D}, we write $\vec{\theta}=(\vec{\vartheta},\phi_{\mathrm{ref}})$, where $\vec{\vartheta}$ contains the 20 parameters inferred directly by the flow: 14 standard GW parameters and six PN deformations.
Here, $\phi_{\mathrm{ref}}$ denotes the orbital phase at the reference frequency $f_{\mathrm{ref}}=20\,\mathrm{Hz}$.
The flow defines the transformation
\begin{equation}
\vec{\vartheta} = f_\lambda(\mathbf{u}; d),
\label{eq:flow_map}
\end{equation}
where $\mathbf{u}$ is drawn from the base distribution $p(\mathbf{u})$, $d$ denotes the data, and $\lambda$ denotes the network parameters.
By the change-of-variables formula, the learned conditional density is
\begin{equation}
q_\lambda(\vec{\vartheta} \mid d) = p\bigl(f_\lambda^{-1}(\vec{\vartheta}; d)\bigr)\,
\left|\det J_{f_\lambda^{-1}}(\vec{\vartheta}; d)\right|,
\label{eq:npe_density}
\end{equation}
where $J_{f_\lambda^{-1}}$ is the Jacobian of the inverse transformation.

For both events, the network is trained to infer this 20-dimensional, phase-marginalized posterior; the reference orbital phase is sampled synthetically at the importance-sampling stage.
Our analyses use H1 and L1 data in $8\,\mathrm{s}$ segments sampled at $4096\,\mathrm{Hz}$, with a frequency range of $20$--$1024\,\mathrm{Hz}$ and resolution $\Delta f=0.125\,\mathrm{Hz}$.

Training minimizes the data-averaged Kullback--Leibler divergence from the posterior to the flow density. Equivalently, it minimizes the negative log-likelihood over training pairs drawn from the joint distribution $p(\vec{\vartheta}, d \mid H)$:
\begin{equation}
\begin{split}
\mathcal{J}_{\mathrm{NPE}}(\lambda)
&= \mathbb{E}_{p(\vec{\vartheta}, d \mid H)}\bigl[-\log q_\lambda(\vec{\vartheta} \mid d)\bigr]\\
&\approx -\frac{1}{N}\sum_{i=1}^{N} \log q_\lambda\bigl(\vec{\vartheta}^{(i)} \mid d^{(i)}\bigr),
\end{split}
\label{eq:npe_loss}
\end{equation}
where $\{\vec{\vartheta}^{(i)}, d^{(i)}\}$ are the simulated parameter--data pairs.
Reusing the trained network across events within its training domain amortizes the training cost. Proposal generation is fast, while the total inference time also includes synthetic phase sampling and likelihood evaluations for importance weighting.

The learned density $q_\lambda(\vec{\vartheta}\mid d)$ approximates the phase-marginalized posterior.
In DINGO-IS~\cite{Dax:2022pxd}, we augment each flow sample with a phase drawn from a conditional proposal $q_\phi(\phi_{\mathrm{ref}}\mid\vec{\vartheta},d)$.
The resulting 21-dimensional proposal density used for importance sampling is
\begin{equation}
q(\vec{\theta}\mid d)
= q_\lambda(\vec{\vartheta}\mid d)\,
q_\phi(\phi_{\mathrm{ref}}\mid\vec{\vartheta},d),
\label{eq:full_proposal}
\end{equation}
which includes the density contribution from synthetic phase sampling.
For $n$ samples $\vec{\theta}_i\sim q(\vec{\theta}\mid d)$, the importance weights are
\begin{equation}
w_i = \frac{\mathcal{L}(d \mid \vec{\theta}_i, H)\,\pi(\vec{\theta}_i \mid H)}{q(\vec{\theta}_i \mid d)}.
\label{eq:is_weights}
\end{equation}
We train with a prior uniform in luminosity distance $d_L$ to increase the representation of nearby signals. For the observed events, DINGO-IS and \texttt{bilby-tgr} adopt a prior uniform in comoving volume over the same distance bounds.
Provided that the proposal covers the support of the target posterior, self-normalized importance-weighted estimates of integrable posterior quantities converge to their expectations under Eq.~\eqref{eq:bayesian_posterior} as the sample size increases.
The quality of the proposal is quantified by the Kish effective sample size (ESS),
\begin{equation}
n_{\mathrm{eff}} = \frac{\bigl(\sum_i w_i\bigr)^2}{\sum_i w_i^2},
\label{eq:n_eff}
\end{equation}
and the corresponding efficiency $\epsilon = n_{\mathrm{eff}}/n$.
The same weights provide an unbiased estimate of the Bayesian evidence,
\begin{equation}
Z = p(d \mid H) = \frac{1}{n}\sum_{i=1}^{n} w_i.
\label{eq:evidence}
\end{equation}

The event-specific training data, priors, network settings, and calibration tests are described in Appendix~\ref{app:training_settings}. Training is monitored on a held-out validation set using the validation counterpart of Eq.~\eqref{eq:npe_loss}.

The importance-weighted samples representing $p(\vec{\theta}_{\mathrm{GR}}, \vec{\theta}_{\mathrm{D}} \mid d, H)$ serve as the input to the PCA described next.

\subsection{Principal component analysis}
The fractional deformation parameters of Eq.~\eqref{eq:theta_D} are generally strongly correlated. Multiparameter tests directly in terms of the $\delta\hat{\phi}_b$ can therefore yield broad, degenerate posteriors that carry little information about individual deviations~\cite{Mahapatra:2025cwk}.
We address this issue by applying PCA to the marginalized deviation-parameter space to identify the best-constrained linear combinations of the original parameters~\cite{Mahapatra:2025cwk}.

Concretely, under a hypothesis $H$ describing the signal model, Bayes' theorem gives the joint posterior of the GR parameters $\vec{\theta}_{\mathrm{GR}}$ and the deviation parameters $\vec{\theta}_{\mathrm{D}}$ as
\begin{equation}
\begin{split}
p(\{\vec{\theta}_{\mathrm{GR}}, \vec{\theta}_{\mathrm{D}}\} \mid d, H)
&= \frac{\mathcal{L}(d \mid \{\vec{\theta}_{\mathrm{GR}}, \vec{\theta}_{\mathrm{D}}\}, H)}{Z(d \mid H)}\\
&\quad\times \pi(\{\vec{\theta}_{\mathrm{GR}}, \vec{\theta}_{\mathrm{D}}\} \mid H).
\end{split}
\label{eq:bayesian_posterior}
\end{equation}

Since our focus is on the fractional deformation parameters, we marginalize Eq.~\eqref{eq:bayesian_posterior} over the GR parameters to obtain the joint posterior of $\vec{\theta}_{\mathrm{D}}$,
\begin{equation}
p(\vec{\theta}_{\mathrm{D}} \mid d, H) = \int p(\{\vec{\theta}_{\mathrm{GR}}, \vec{\theta}_{\mathrm{D}}\} \mid d, H)\, d\vec{\theta}_{\mathrm{GR}}.
\label{eq:marginalization}
\end{equation}
In our framework, this marginalized posterior is represented by the posterior samples produced by DINGO NPE with importance-sampling correction in Sec.~\ref{subsec:npe}. Each sample $\vec{\theta}_i$ carries the importance weight $w_i$ of Eq.~\eqref{eq:is_weights}, and the resulting weighted sample set provides a Monte Carlo approximation to Eq.~\eqref{eq:marginalization}.

We estimate the covariance matrix of the deviation parameters from these posterior samples.
The angular brackets denote importance-weighted averages over the marginalized posterior,
\begin{equation}
\langle x \rangle = \frac{\sum_i w_i\, x_i}{\sum_i w_i},
\label{eq:weighted_average}
\end{equation}
which reduces to the ordinary sample average when all weights are equal.
The covariance matrix of the PN deformation parameters is then defined as
\begin{equation}
C_{ab} = \Big\langle \big(\delta\hat{\phi}_a - \langle \delta\hat{\phi}_a \rangle\big)\big(\delta\hat{\phi}_b - \langle \delta\hat{\phi}_b \rangle\big) \Big\rangle.
\label{eq:covariance_matrix}
\end{equation}

Being real and symmetric, $C$ is diagonalized by an orthogonal transformation,
\begin{equation}
C = U S U^{\mathrm{T}},
\label{eq:eigen_decomposition}
\end{equation}
where $S = \mathrm{diag}(S_1, S_2, \ldots, S_6)$ is the diagonal matrix of eigenvalues and $U$ is the orthogonal matrix whose columns are the corresponding eigenvectors.

The eigenvectors define the new basis.
Writing the components of eigenvector $i$ as $v_b^{(i)} \equiv U_{bi}$, we obtain the PCA parameters by rotating the original deviation parameters,
\begin{equation}
\delta\hat{\phi}_{\mathrm{PCA}}^{(i)} = \sum_b v_b^{(i)}\, \delta\hat{\phi}_b.
\label{eq:pca_transformation}
\end{equation}
For the posterior covariance used to construct the basis, the PCA parameters are uncorrelated and their variances equal the corresponding eigenvalues,
\begin{equation}
\mathrm{Cov}\bigl(\delta\hat{\phi}_{\mathrm{PCA}}^{(i)}, \delta\hat{\phi}_{\mathrm{PCA}}^{(j)}\bigr) = S_i\, \delta_{ij}.
\label{eq:pca_decorrelation}
\end{equation}
When a different posterior is projected onto this fixed basis, its covariance need not be diagonal. Such projections allow constraints to be compared along the same linear combinations, with dimensionality reduction achieved by retaining only the well-constrained modes.
We order the eigenvalues in ascending order, so the mode with the smallest posterior variance, $i=1$, is referred to as the leading PCA mode.
We denote the leading and subleading PCA modes by PC1 and PC2, respectively.
To determine which modes are informed by the data rather than predominantly constrained by the prior, we compare their posterior and prior distributions.
For this comparison, we follow Ref.~\cite{Mahapatra:2025cwk} and adopt a Jensen--Shannon divergence (JSD) criterion.
For each PCA mode, we compute the JSD between its posterior $p_{\mathrm{post}}$ and prior $p_{\mathrm{prior}}$,
\begin{equation}
\begin{aligned}
D_{\mathrm{JS}}
&= \frac{1}{2}\left[D_{\mathrm{KL}}(p_{\mathrm{post}} \| m) + D_{\mathrm{KL}}(p_{\mathrm{prior}} \| m)\right],\\
m &= \frac{p_{\mathrm{post}} + p_{\mathrm{prior}}}{2}.
\end{aligned}
\label{eq:js_divergence}
\end{equation}
Here $D_{\mathrm{KL}}$ uses base-2 logarithms, so $D_{\mathrm{JS}}$ is measured in bits.
A JSD close to zero indicates that the posterior of a mode is almost indistinguishable from its prior. The data then carry essentially no information about that mode, whose constraint is dominated by the prior; we truncate and discard such modes.
We retain only the modes that satisfy $D_{\mathrm{JS}}\bigl(p_{\mathrm{post}},p_{\mathrm{prior}}\bigr) > 0.1$ bits,
with the threshold chosen following Ref.~\cite{Mahapatra:2025cwk}.

Applying this criterion to the six-dimensional PN-deformation posteriors, we find that only the leading and subleading modes exceed the 0.1-bit threshold and retain these two modes for subsequent analysis. This is consistent with Fisher-matrix forecasts for next-generation ground-based detectors, in which the two best-constrained modes capture more than $99\%$ of the information~\cite{Datta:2022izc}; our selection is based on JSD rather than an assumed information fraction.

PCA has an inherent sign ambiguity: an eigenvector and its negative diagonalize $C$ identically.
We fix this ambiguity by adopting the convention that the element of each eigenvector with the largest absolute value is positive.
This fixes the orientation of each PCA axis without changing its statistical properties; the linear combinations may still differ across injections and events.

The PCA parameters inherit the null-test character of the deformation parameters: since all $\delta\hat{\phi}_b$ vanish under GR, all $\delta\hat{\phi}_{\mathrm{PCA}}^{(i)}$ vanish as well, and the test of GR reduces to checking whether the posterior of each PCA mode is consistent with zero.
We quantify this with the GR quantile
\begin{equation}
\mathcal{Q}_{\mathrm{GR}} = P\bigl(\delta\hat{\phi}_{\mathrm{PCA}}^{(i)} < 0\bigr),
\label{eq:Q_GR}
\end{equation}
The GR quantile is close to $0.5$ when the posterior is symmetric around zero and approaches $0$ or $1$ when the GR value lies in a tail. We also use the GR $z$-score $z_{\mathrm{GR}}$, defined as the number of standard deviations by which the GR value (zero) is displaced from the posterior median. We retain the sign, with positive scores corresponding to negative posterior medians and vice versa.
A PCA mode is deemed consistent with GR when zero lies within its $90\%$ credible interval; we report $\mathcal{Q}_{\mathrm{GR}}$ and $z_{\mathrm{GR}}$ for the leading modes in Sec.~\ref{sec:results}.

Denoting the leading PCA mode by $\alpha_1\equiv\delta\hat{\phi}_{\mathrm{PCA}}^{(1)}$, we assess whether the GR value lies in a low-density region of the posterior using the highest-posterior-density (HPD) credibility statistic~\cite{Crisostomi:2023tle}
\begin{equation}\label{eq:T_HPD}
    T_{\mathrm{HPD}} = \int_{p(\alpha_1\mid d) > p(0\mid d)} p(\alpha_1\mid d)\,d\alpha_1,
\end{equation}
which is the posterior probability mass at densities higher than that at GR. This definition accommodates multimodal posteriors and does not encode the sign of the deviation. Here $T_{\mathrm{HPD}}=0$ when GR lies at the posterior maximum and approaches 1 when GR lies in a very low-density region. We compare the observed value with an empirical distribution from GR injections to assess how often equal or larger values arise under the adopted noise model.

In this work, the prior and inference ranges for each PN deviation parameter $\delta\hat{\phi}_b$ are both $[-5,5]$. We choose this range to balance the extent of the deformation parameter space against the difficulty of training.

\section{Results}
\label{sec:results}
\begin{table*}[!htbp]
\caption{GR injection configurations and PCA results. The chirp masses $M_c$ are given in the detector frame. N denotes nonprecessing spins with $(a_1,a_2)=(0.01,0.01)$ and all four spin angles zero. P denotes precessing spins with $(a_1,a_2)=(0.3002,0.2002)$, $(\theta_1,\theta_2)=(1.5375,1.5208)$, and $(\phi_{12},\phi_{JL})=(\pi/2,1.0)$. No.~9 denotes the GW150914-like reference configuration.}
\label{results-16inj}
\renewcommand{\arraystretch}{1.60}
\begin{ruledtabular}
\begin{tabular}{ccccccccc}
No. & $M_c$ ($M_\odot$) & $q$ & Spin & $d_L$ (Mpc) & \multicolumn{2}{c}{$\delta\hat{\phi}_{\mathrm{PCA}}^{(1)}$} & \multicolumn{2}{c}{$\delta\hat{\phi}_{\mathrm{PCA}}^{(2)}$} \\
\cline{6-7} \cline{8-9}
 &  &  &  &  & Median and 90\% errors & $\mathcal{Q}_{\mathrm{GR}}$ ($z_{\mathrm{GR}}$) & Median and 90\% errors & $\mathcal{Q}_{\mathrm{GR}}$ ($z_{\mathrm{GR}}$) \\
\hline
1 & 27.0 & 0.81 & N & 440 & $-0.03^{+0.16}_{-0.18}$ & 0.595 (0.24) & $-0.01^{+1.58}_{-1.50}$ & 0.502 (0.01) \\
2 &  &  &  & 300 & $-0.01^{+0.12}_{-0.13}$ & 0.531 (0.08) & $-0.05^{+1.09}_{-1.10}$ & 0.534 (0.08) \\
3 & 27.0 & 0.81 & P & 440 & $-0.03^{+0.16}_{-0.18}$ & 0.628 (0.32) & $+0.03^{+1.56}_{-1.52}$ & 0.487 ($-0.03$) \\
4 &  &  &  & 300 & $-0.02^{+0.12}_{-0.13}$ & 0.584 (0.21) & $-0.07^{+1.09}_{-1.14}$ & 0.541 (0.10) \\
5 & 27.0 & 0.45 & N & 440 & $-0.13^{+0.23}_{-0.27}$ & 0.809 (0.83) & $-0.71^{+2.04}_{-2.15}$ & 0.722 (0.55) \\
6 &  &  &  & 300 & $-0.05^{+0.15}_{-0.18}$ & 0.697 (0.49) & $+0.46^{+1.50}_{-1.41}$ & 0.310 ($-0.52$) \\
7 & 27.0 & 0.45 & P & 440 & $-0.09^{+0.23}_{-0.27}$ & 0.749 (0.62) & $-0.47^{+2.06}_{-2.18}$ & 0.637 (0.36) \\
8 &  &  &  & 300 & $-0.03^{+0.15}_{-0.19}$ & 0.625 (0.31) & $+0.32^{+1.63}_{-1.59}$ & 0.371 ($-0.33$) \\
9 & 31.2 & 0.81 & N & 440 & $-0.01^{+0.16}_{-0.16}$ & 0.541 (0.10) & $+0.12^{+2.02}_{-1.84}$ & 0.459 ($-0.10$) \\
10 &  &  &  & 300 & $-0.00^{+0.12}_{-0.11}$ & 0.513 (0.03) & $+0.08^{+1.43}_{-1.27}$ & 0.458 ($-0.10$) \\
11 & 31.2 & 0.81 & P & 440 & $-0.01^{+0.16}_{-0.16}$ & 0.525 (0.06) & $+0.25^{+2.00}_{-1.84}$ & 0.410 ($-0.21$) \\
12 &  &  &  & 300 & $+0.00^{+0.11}_{-0.11}$ & 0.489 ($-0.03$) & $+0.23^{+1.41}_{-1.28}$ & 0.383 ($-0.28$) \\
13 & 31.2 & 0.45 & N & 440 & $-0.06^{+0.19}_{-0.19}$ & 0.684 (0.53) & $+0.13^{+2.39}_{-2.36}$ & 0.470 ($-0.09$) \\
14 &  &  &  & 300 & $-0.03^{+0.14}_{-0.15}$ & 0.666 (0.39) & $+0.02^{+1.68}_{-1.36}$ & 0.478 ($-0.03$) \\
15 & 31.2 & 0.45 & P & 440 & $+0.01^{+0.21}_{-0.19}$ & 0.466 ($-0.08$) & $-0.09^{+2.23}_{-1.91}$ & 0.525 (0.07) \\
16 &  &  &  & 300 & $+0.02^{+0.21}_{-0.20}$ & 0.438 ($-0.18$) & $+0.04^{+1.46}_{-1.49}$ & 0.493 ($-0.04$) \\
\end{tabular}
\end{ruledtabular}
\end{table*}
\begin{table*}[!htbp]
\caption{BGR injection configurations and PCA results for detector-frame chirp mass $M_c=31.2\,M_\odot$, mass ratio $q=0.81$, and the N/P spin configurations defined in Table~\ref{results-16inj}. The entries of $\{\delta\hat{\phi}_b\}$ are ordered as $(\delta\hat{\phi}_3,\delta\hat{\phi}_4,\delta\hat{\phi}_{5\ell},\delta\hat{\phi}_6,\delta\hat{\phi}_{6\ell},\delta\hat{\phi}_7)$; uniform patterns are abbreviated as $\delta\hat{\phi}_b=c$. Each pattern is tested for both spin configurations at two distances.}
\label{results-12inj}
\setlength{\tabcolsep}{1pt}
\renewcommand{\arraystretch}{1.60}
\begin{ruledtabular}
\begin{tabular}{clcccccc}
No. & $\{\delta\hat{\phi}_b\}$ & Spin & $d_L$ (Mpc) & \multicolumn{2}{c}{$\delta\hat{\phi}_{\mathrm{PCA}}^{(1)}$} & \multicolumn{2}{c}{$\delta\hat{\phi}_{\mathrm{PCA}}^{(2)}$} \\
\cline{5-6} \cline{7-8}
 &  &  &  & Median and 90\% errors & $\mathcal{Q}_{\mathrm{GR}}$ ($z_{\mathrm{GR}}$) & Median and 90\% errors & $\mathcal{Q}_{\mathrm{GR}}$ ($z_{\mathrm{GR}}$) \\
\hline
1 & $\delta\hat{\phi}_b=0.1$ & N & 440 & $+0.08^{+0.15}_{-0.16}$ & 0.192 ($-0.89$) & $+0.04^{+1.97}_{-1.84}$ & 0.489 ($-0.03$) \\
2 &  &  & 300 & $+0.09^{+0.11}_{-0.12}$ & 0.087 ($-1.34$) & $+0.04^{+1.37}_{-1.29}$ & 0.479 ($-0.05$) \\
3 &  & P & 440 & $+0.08^{+0.16}_{-0.16}$ & 0.191 ($-0.88$) & $+0.22^{+1.97}_{-1.87}$ & 0.426 ($-0.19$) \\
4 &  &  & 300 & $+0.09^{+0.11}_{-0.11}$ & 0.088 ($-1.35$) & $+0.17^{+1.41}_{-1.25}$ & 0.426 ($-0.21$) \\
\hline
5 & $\delta\hat{\phi}_b=0.5$ & N & 440 & $+0.50^{+0.11}_{-0.11}$ & 0.000 ($-7.35$) & $+0.08^{+1.88}_{-1.99}$ & 0.473 ($-0.07$) \\
6 &  &  & 300 & $+0.51^{+0.07}_{-0.07}$ & 0.000 ($-11.14$) & $+0.13^{+1.26}_{-1.41}$ & 0.435 ($-0.16$) \\
7 &  & P & 440 & $+0.50^{+0.11}_{-0.12}$ & 0.000 ($-7.30$) & $+0.34^{+1.86}_{-1.97}$ & 0.386 ($-0.29$) \\
8 &  &  & 300 & $+0.50^{+0.08}_{-0.08}$ & 0.000 ($-10.69$) & $+0.39^{+1.28}_{-1.42}$ & 0.318 ($-0.47$) \\
\hline
9 & $\{0.3,0.4,0.5,0.6,0.7,0.8\}$ & N & 440 & $+0.33^{+0.13}_{-0.14}$ & 0.000 ($-3.98$) & $-0.02^{+1.89}_{-1.96}$ & 0.508 (0.02) \\
10 &  &  & 300 & $+0.34^{+0.10}_{-0.10}$ & 0.000 ($-5.72$) & $-0.07^{+1.33}_{-1.34}$ & 0.531 (0.08) \\
11 &  & P & 440 & $+0.32^{+0.14}_{-0.14}$ & 0.000 ($-3.83$) & $-0.18^{+2.00}_{-1.92}$ & 0.560 (0.15) \\
12 &  &  & 300 & $+0.33^{+0.10}_{-0.10}$ & 0.000 ($-5.54$) & $+0.13^{+1.30}_{-1.33}$ & 0.441 ($-0.16$) \\
\end{tabular}
\end{ruledtabular}
\end{table*}

In this section, we assess posterior recovery in the PCA parameter space.
We first analyze GR injections and BGR injections with nonzero PN deformation parameters to examine consistency with GR and the response to injected deviations at different deformation strengths and SNRs.
Next, we analyze GW150914 and GW250114 and compute their PCA modes. To assess consistency, we independently repeat the inference and PCA calculations with \texttt{bilby-tgr} and compare the results with previous PCA analyses of GW150914~\cite{Kumar:2025jio} and GW250114~\cite{LIGOScientific:2025wao}.
Finally, we perform a large-scale injection--recovery study using a fixed GW150914-like GR signal and 1000 independent noise realizations to assess the apparent deviation observed in GW150914.

\begin{figure*}[!htbp]
	\centering
	\includegraphics[width=0.95\linewidth]{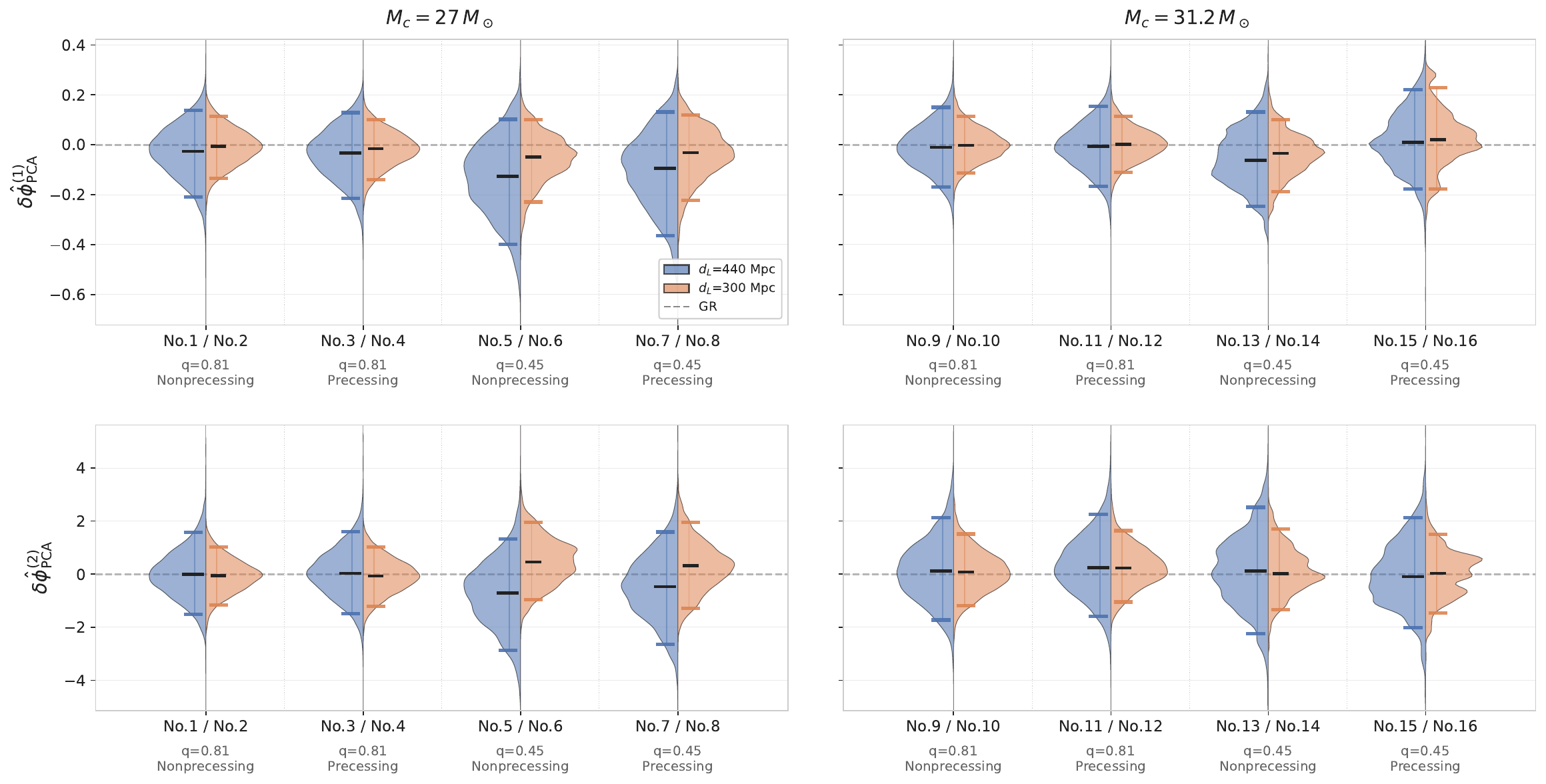}
	\caption{Posteriors of the two best-constrained PCA modes for 16 zero-noise GR injections. The PCA basis is constructed separately from the posterior of each injection. The upper and lower rows show $\delta\hat{\phi}_{\mathrm{PCA}}^{(1)}$ and $\delta\hat{\phi}_{\mathrm{PCA}}^{(2)}$, while the left and right columns correspond to detector-frame chirp masses $M_c=27\,M_\odot$ and $31.2\,M_\odot$. Each split violin compares $d_L=440$ Mpc (blue, left) with $300$ Mpc (orange, right) for the indicated mass ratio and spin configuration. The nonprecessing and precessing spin configurations are defined in Table~\ref{results-16inj}. Black bars mark posterior medians, colored bars show 90\% equal-tailed credible intervals, and the dashed line denotes the GR value.}
	\label{fig:pca_gr16_zeronoise_250k_big_figure}
\end{figure*}
\begin{figure*}[!htbp]
	\centering
	\includegraphics[width=0.95\linewidth]{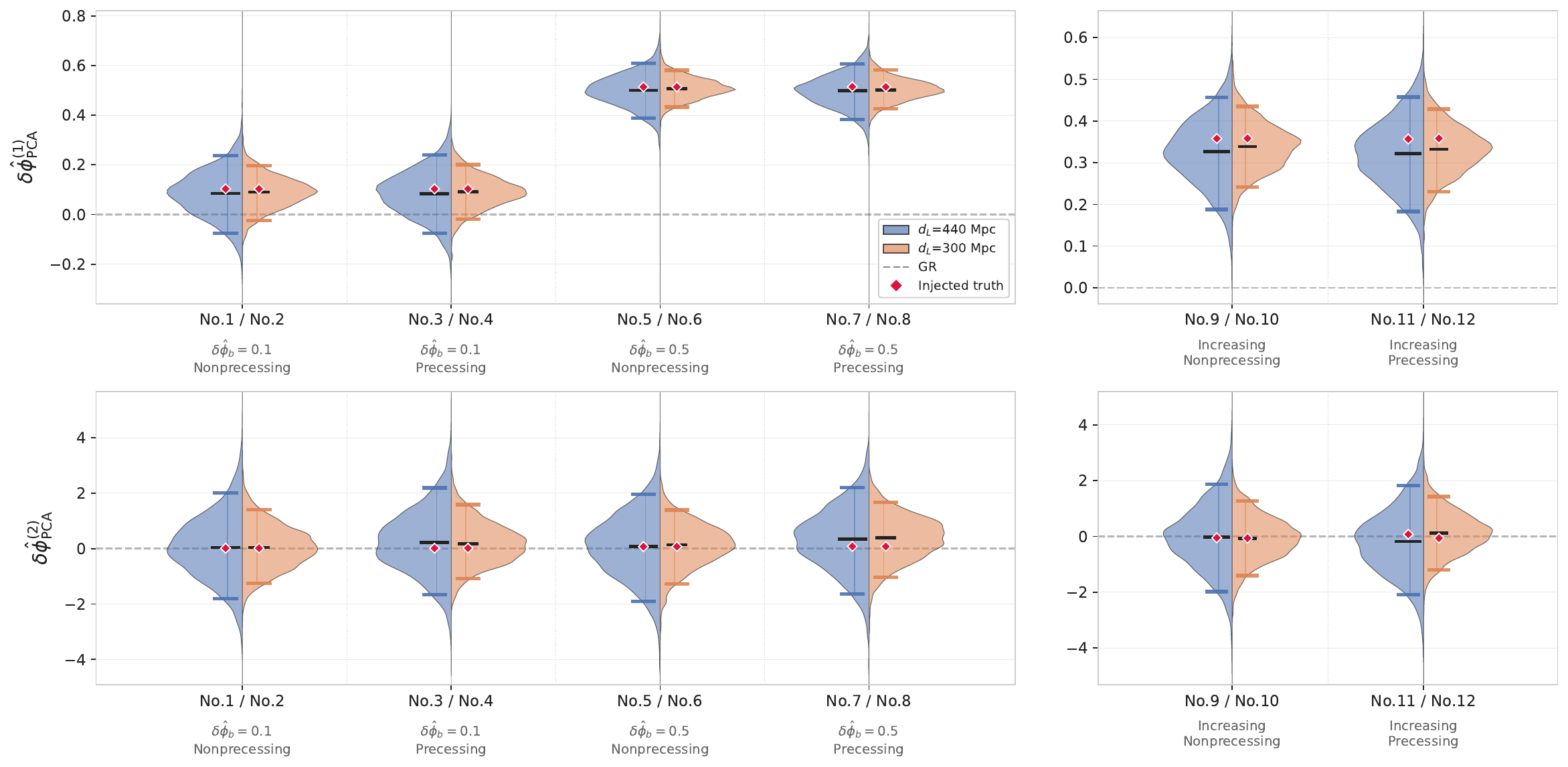}
	\caption{Posteriors of the two best-constrained PCA modes for 12 zero-noise BGR injections with GW150914-like binary parameters. The PCA basis is constructed separately from the posterior of each injection. The upper and lower rows show $\delta\hat{\phi}_{\mathrm{PCA}}^{(1)}$ and $\delta\hat{\phi}_{\mathrm{PCA}}^{(2)}$ for the deformation patterns and spin configurations labeled below each violin; the increasing pattern is displayed separately on the right. Each split violin compares $d_L=440$ Mpc (blue, left) with $300$ Mpc (orange, right). Uniform patterns assign equal values to all six PN deformation parameters. The increasing pattern follows the parameter order specified in Table~\ref{results-12inj}, which lists all injection vectors. Red diamonds denote the injected values, black bars mark posterior medians, colored bars show 90\% equal-tailed credible intervals, and the dashed line denotes GR.}
	\label{fig:pca_bgr12_zeronoise_big_figure}
\end{figure*}
\subsection{Recovery validation: GR and BGR injections}
To assess posterior recovery with our pipeline, we perform two sets of zero-noise injection--recovery tests, following the general approach of previous PCA-based studies using conventional samplers~\cite{Mahapatra:2025cwk}.
The GR injections test whether the recovered PCA posteriors remain consistent with zero, while the BGR injections probe the response to nonzero PN deformations of different strengths. For each injection, we construct the PCA basis from its importance-weighted posterior and project the injected deformation vector onto the same basis. Thus, PC1 and PC2 need not represent identical PN combinations across injections.

We consider 16 GR injections following the grouping scheme of Ref.~\cite{Mahapatra:2025cwk}, with the configurations listed in Table~\ref{results-16inj}.
The injections are grouped by detector-frame chirp mass, with $M_c=27.0\,M_\odot$ and $31.2\,M_\odot$, and span different mass ratios, precession states, and luminosity distances.
Injection 9 serves as the GW150914-like reference configuration.

For the BGR injections, we consider uniform deformations $\delta\hat{\phi}_b=0.1$ and $0.5$, together with the increasing pattern listed in Table~\ref{results-12inj}.
Each pattern is tested for both N/P spin configurations at two luminosity distances, corresponding to different SNRs.
The chirp mass and mass ratio are fixed to the GW150914-like values listed in Table~\ref{results-12inj}.

Figure~\ref{fig:pca_gr16_zeronoise_250k_big_figure} shows the posteriors of the two retained PCA modes for the GR injections.
The nearer, higher-SNR configurations generally yield tighter constraints, while the more unequal-mass configurations generally yield broader posteriors.
The N/P comparison keeps $\chi_{\mathrm{eff}}\simeq0.01$ approximately fixed while changing $\chi_p$ from zero to about $0.30$.
For all 16 injections, zero lies within the $90\%$ credible intervals of both modes.
The BGR injection configurations and recovered PCA posteriors are shown in Table~\ref{results-12inj} and Fig.~\ref{fig:pca_bgr12_zeronoise_big_figure}, respectively.
The leading mode is more tightly constrained than the subleading mode.
For the uniform positive injections, its displacement from GR increases with the injected deformation amplitude.
All 12 injected PC1 values lie within the corresponding $90\%$ credible intervals; GR is excluded from these intervals for the $0.5$ and increasing patterns.
The tested patterns are specified in the original PN parameters and yield small PC2 offsets relative to the posterior uncertainties.
The injected PC2 values lie close to the posterior medians, with the $90\%$ credible intervals including both the injected values and zero.
This behavior is also seen in Fig.~6 of Ref.~\cite{Mahapatra:2025cwk}.
Dedicated injections testing recovery of a resolvable nonzero PC2 displacement are not included.
We project the three deformation patterns in Table~\ref{results-12inj} onto the GW150914 \texttt{bilby-tgr} PC1 direction in Table~\ref{tab:pca_coefficients}. In each case, the 1.5PN term gives the largest absolute contribution.

These zero-noise tests support the recovery of GR-consistent PCA posteriors in the GW150914-like parameter region considered here.

\subsection{Real-event application}
\label{subsec:real_events}
We next apply the pipeline to GW150914 and GW250114 and report the two PCA modes selected by the JSD criterion.

For each event, we compare the DINGO-IS posteriors with those from an independent \texttt{bilby-tgr}~\cite{ashton_2025_15676285} analysis using the same waveform model and the prior bounds used for DINGO training.
For both events, the DINGO-IS likelihood uses noise power spectral densities (PSDs) estimated from $1024\,\mathrm{s}$ of strain data per detector.

\subsubsection*{GW150914}\label{subsubsec:GW150914}
GW150914, the first GW event observed by ground-based interferometers, serves as a well-established benchmark for testing GW inference methods. We therefore apply our approach to the publicly available GW150914 strain data from the Gravitational Wave Open Science Center (GWOSC)~\cite{LIGOScientific:2016GW150914} and perform Bayesian inference of the corresponding PCA parameters.

\begin{figure}[tb]
	\centering
	\includegraphics[width=\linewidth]{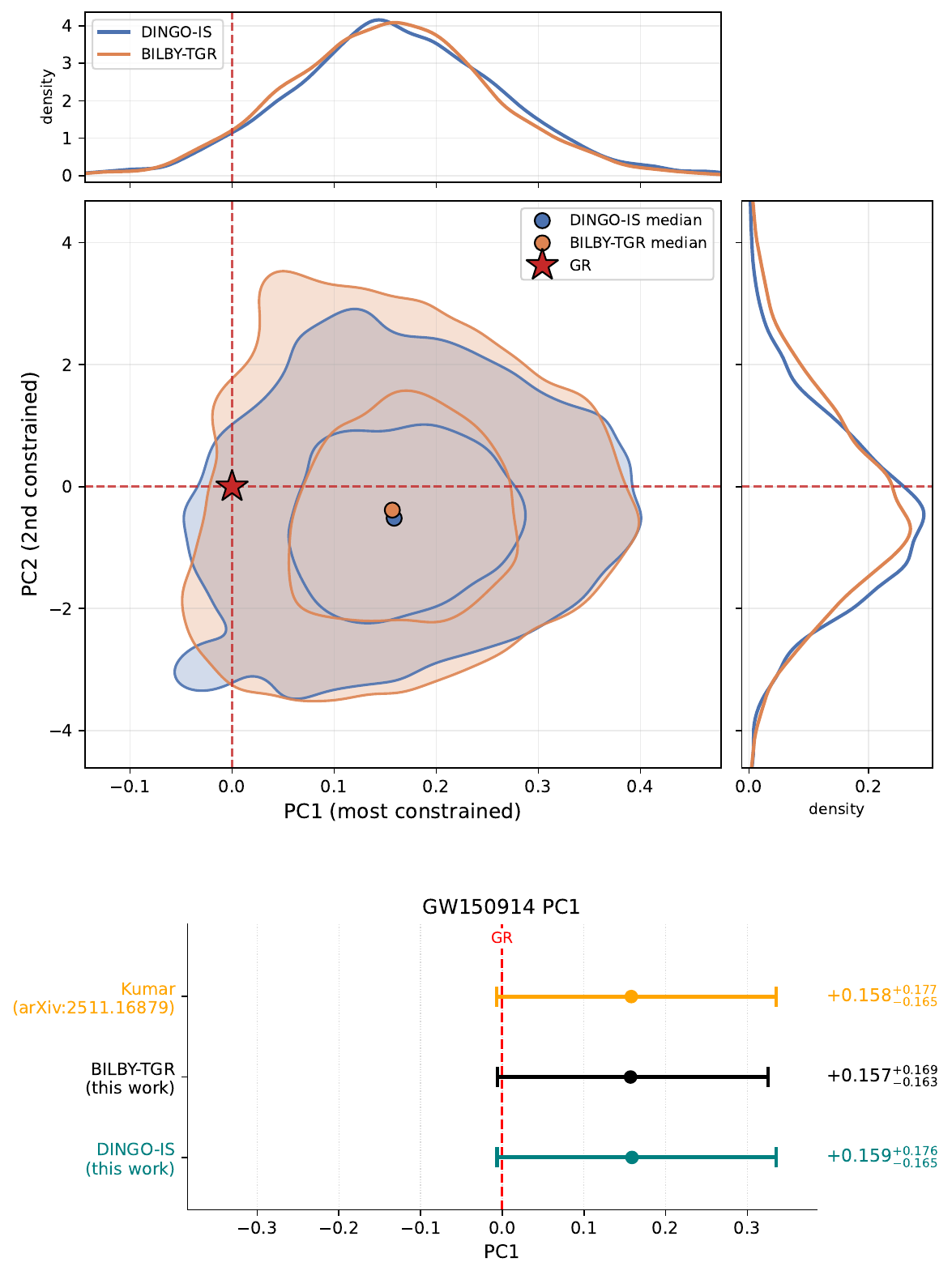}
	\caption{PCA posteriors for GW150914 in the common eigenbasis of its \texttt{bilby-tgr} posterior. The joint distribution and marginal densities of the leading (PC1) and subleading (PC2) modes compare DINGO-IS (blue) with \texttt{bilby-tgr} (orange). Circles mark the posterior medians, the red star denotes GR, and shaded contours show the 50\% and 90\% 2D HPD credible regions. The lower panel compares the PC1 medians and 90\% equal-tailed credible intervals from both methods with the PCA result of Ref.~\cite{Kumar:2025jio}. Both inference methods give a PC1 median near $0.16$, with the GR value inside the 90\% interval near its lower edge.}
	\label{fig:gw150914_pca_position}
\end{figure}

We analyze GW150914 using an NPE model trained on three million simulated BBH signals spanning the parameter region containing the event posterior; the prior ranges are listed in Table~\ref{tab:GW150914priors}.
We train with O1 PSDs and allow precessing spins.
\begin{table}[!tb]
\centering
\caption{Training priors for the parameters directly inferred by the 20-dimensional GW150914 model. Masses are given in the detector frame (i.e., redshifted masses); $\mathcal{U}$ denotes a uniform distribution.}
\label{tab:GW150914priors}
\begin{tabular}{c c l}
\toprule
Parameter & Unit & Prior \\
\midrule
$M_c$ & $M_\odot$ & $\mathcal{U}(25,\,35)$ \\
$q$ & $-$ & $\mathcal{U}(0.25,\,1)$ \\
$a_1$ & $-$ & $\mathcal{U}(0,\,0.9)$ \\
$a_2$ & $-$ & $\mathcal{U}(0,\,0.9)$ \\
$\theta_1$ & rad & $\sin(0,\,\pi)$ \\
$\theta_2$ & rad & $\sin(0,\,\pi)$ \\
$\phi_{12}$ & rad & $\mathcal{U}(0,\,2\pi)$ \\
$\phi_{JL}$ & rad & $\mathcal{U}(0,\,2\pi)$ \\
$\theta_{\rm JN}$ & rad & $\sin(0,\,\pi)$ \\
$d_L$ & Mpc & $\mathcal{U}(150,\,800)$ \\
$t_c$ & s & $\mathcal{U}(-0.1,\,0.1)$ \\
$\delta$ & rad & $\cos(-\pi/2,\,\pi/2)$ \\
$\alpha$ & rad & $\mathcal{U}(0,\,2\pi)$ \\
$\psi$ & rad & $\mathcal{U}(0,\,\pi)$ \\
$\delta\hat{\phi}_b$ & $-$ & $\mathcal{U}(-5,\,5)$ \\
\bottomrule
\end{tabular}
\end{table}
Additional details of the training procedure and calibration, including the probability--probability (P--P) calibration plot, are provided in Appendix~\ref{app:training_settings}.
For the event analysis, we use the importance-sampling-corrected output from DINGO-IS.
We additionally perform \texttt{bilby-tgr} inference for GW150914 using the training prior bounds and precomputed O1 noise spectra nearest to the trigger time.
With one NVIDIA RTX 4090 GPU for proposal generation and 128 CPU cores for synthetic phase sampling and importance sampling, the total inference wall-clock time is approximately 38 minutes (74.4 CPU hours for the IS stage). The 350,000 proposal samples yield an ESS of approximately 4231 at an importance-sampling efficiency of $1.21\%$.
The \texttt{bilby-tgr} analysis with \texttt{dynesty} takes 176.9 hours of wall-clock time on 32 CPU cores (498.9 CPU hours) and yields 5371 equal-weight posterior samples.
After marginalizing over the standard GW parameters, we construct a common PCA basis from the \texttt{bilby-tgr} posterior of the six PN deformations. We project both posteriors onto this basis.
Only the first two modes exceed the JSD threshold of $0.1$ bits; their joint posterior is shown in Fig.~\ref{fig:gw150914_pca_position}.

The independently derived DINGO-IS and \texttt{bilby-tgr} eigenvectors differ by $0.378^\circ$ ($\cos\theta=0.99998$) for PC1 and $1.440^\circ$ for PC2.
The PC1 posteriors show a positive offset, with zero remaining within the $90\%$ credible intervals near their lower boundaries.
This result is qualitatively consistent with previous PCA-based multiparameter analyses of GW150914~\cite{Saleem:2021nsb,Kumar:2025jio}.
In particular, Ref.~\cite{Saleem:2021nsb} reported an approximately $2\sigma$ offset of the GR value from the posterior median of the leading PCA parameter, with GR lying just outside its $90\%$ credible interval.

Such an offset does not by itself establish a violation of GR; noise fluctuations and waveform-modeling uncertainties must also be considered~\cite{LIGOScientific:2016lio,LIGOScientific:2019fpa,Mezzasoma:2026wme}.
To assess how unusual the observed offset is under the adopted Gaussian-noise model, we perform a large set of GW150914-like GR injection--recovery tests below.

\subsubsection*{GW250114}
\begin{figure}[!htbp]
	\centering
	\includegraphics[width=\linewidth]{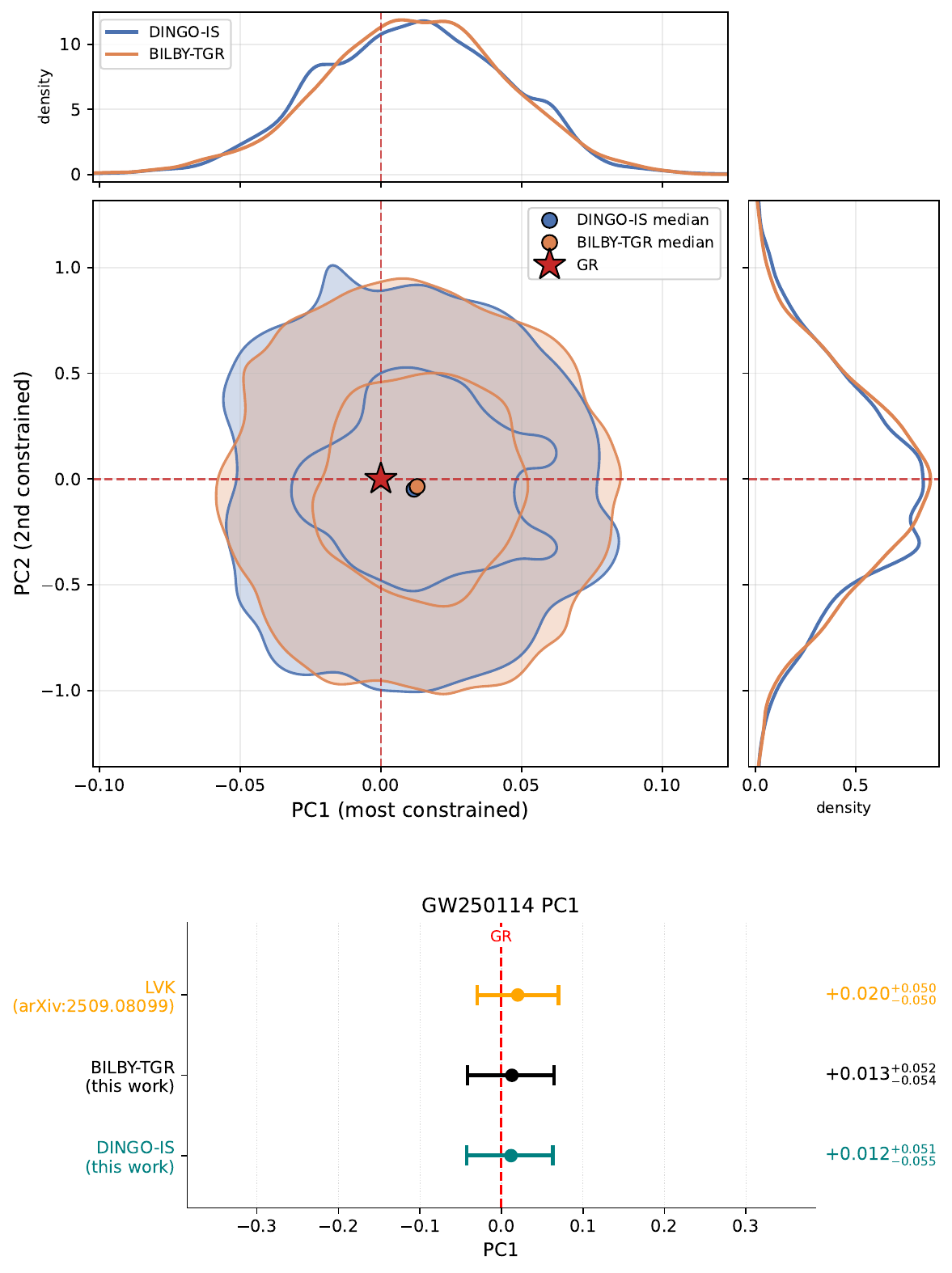}
	\caption{PCA posteriors for GW250114 in the common eigenbasis of its \texttt{bilby-tgr} posterior. The joint distribution and marginal densities of the leading (PC1) and subleading (PC2) modes compare DINGO-IS (blue) with \texttt{bilby-tgr} (orange). Circles mark the posterior medians, the red star denotes GR, and shaded contours show the 50\% and 90\% 2D HPD credible regions. The lower panel compares the PC1 medians and 90\% equal-tailed credible intervals with the LVK result. DINGO-IS gives $0.012^{+0.051}_{-0.055}$, consistent with both \texttt{bilby-tgr} and GR.}
	\label{fig:gw250114_pca_position_with_lvk}
\end{figure}

We also analyze the O4 BBH event GW250114, which has binary parameters comparable to those of GW150914 but an SNR roughly three times as large.
For this higher-SNR event, we generate $3\times10^6$ waveforms over a narrower parameter range for training.
The prior bounds are summarized in Table~\ref{tab:GW250114priors}.
We use the same precessing-spin parametrization for GW250114.
For GW250114, training uses a fixed fiducial O4b PSD for each detector.
The training procedure is described in Appendix~\ref{app:training_settings}.
We also analyze GW250114 with \texttt{bilby-tgr}, using priors consistent with the training ranges, and compare the results of both analyses with the published LVK measurement in Fig.~\ref{fig:gw250114_pca_position_with_lvk}.
\begin{table}[!tb]
\centering
\caption{Training priors for the parameters directly inferred by the 20-dimensional GW250114 model. Masses are given in the detector frame (i.e., redshifted masses); $\mathcal{U}$ denotes a uniform distribution.}
\label{tab:GW250114priors}
\begin{tabular}{c c l}
\toprule
Parameter & Unit & Prior \\
\midrule
$M_c$ & $M_\odot$ & $\mathcal{U}(28,\,34)$ \\
$q$ & $-$ & $\mathcal{U}(0.6,\,1)$ \\
$a_1$ & $-$ & $\mathcal{U}(0,\,0.5)$ \\
$a_2$ & $-$ & $\mathcal{U}(0,\,0.5)$ \\
$\theta_1$ & rad & $\sin(0,\,\pi)$ \\
$\theta_2$ & rad & $\sin(0,\,\pi)$ \\
$\phi_{12}$ & rad & $\mathcal{U}(0,\,2\pi)$ \\
$\phi_{JL}$ & rad & $\mathcal{U}(0,\,2\pi)$ \\
$\theta_{\rm JN}$ & rad & $\sin(0,\,\pi)$ \\
$d_L$ & Mpc & $\mathcal{U}(250,\,600)$ \\
$t_c$ & s & $\mathcal{U}(-0.1,\,0.1)$ \\
$\delta$ & rad & $\cos(-\pi/2,\,\pi/2)$ \\
$\alpha$ & rad & $\mathcal{U}(0,\,2\pi)$ \\
$\psi$ & rad & $\mathcal{U}(0,\,\pi)$ \\
$\delta\hat{\phi}_b$ & $-$ & $\mathcal{U}(-5,\,5)$ \\
\bottomrule
\end{tabular}
\end{table}

For GW250114, we use its \texttt{bilby-tgr} posterior to construct a common PCA basis and project both posteriors onto it.
The independently derived DINGO-IS and \texttt{bilby-tgr} eigenvectors differ by $0.143^\circ$ ($\cos\theta=0.999997$) for PC1 and $0.427^\circ$ for PC2.
For both DINGO-IS and \texttt{bilby-tgr}, the GR value lies within the $90\%$ credible interval of the leading PCA mode.
The two analyses agree closely and are qualitatively consistent with the published LVK conclusion~\cite{LIGOScientific:2025wao}. The PCA basis used here differs from that of the LVK analysis.
For GW250114, DINGO-IS uses one NVIDIA RTX 4090 GPU for proposal generation and 128 CPU cores for synthetic phase sampling and importance sampling. The total wall-clock time is approximately 4.4 h (521.3 CPU hours for the IS stage). The 2.5 million proposal samples yield an ESS of 1037 (efficiency $0.041\%$).
The corresponding \texttt{bilby-tgr} analysis with \texttt{dynesty} takes 237.6 h of wall-clock time on 64 CPU cores (1484.7 CPU hours) and yields 6153 equal-weight posterior samples.
With these resources, DINGO-IS reduces wall-clock time by more than an order of magnitude for both events.

\subsection{GR null tests with GW150914-like injections}
\label{subsec:GW150914like_injections}
\begin{figure}[!htbp]
	\centering
	\includegraphics[width=\linewidth]{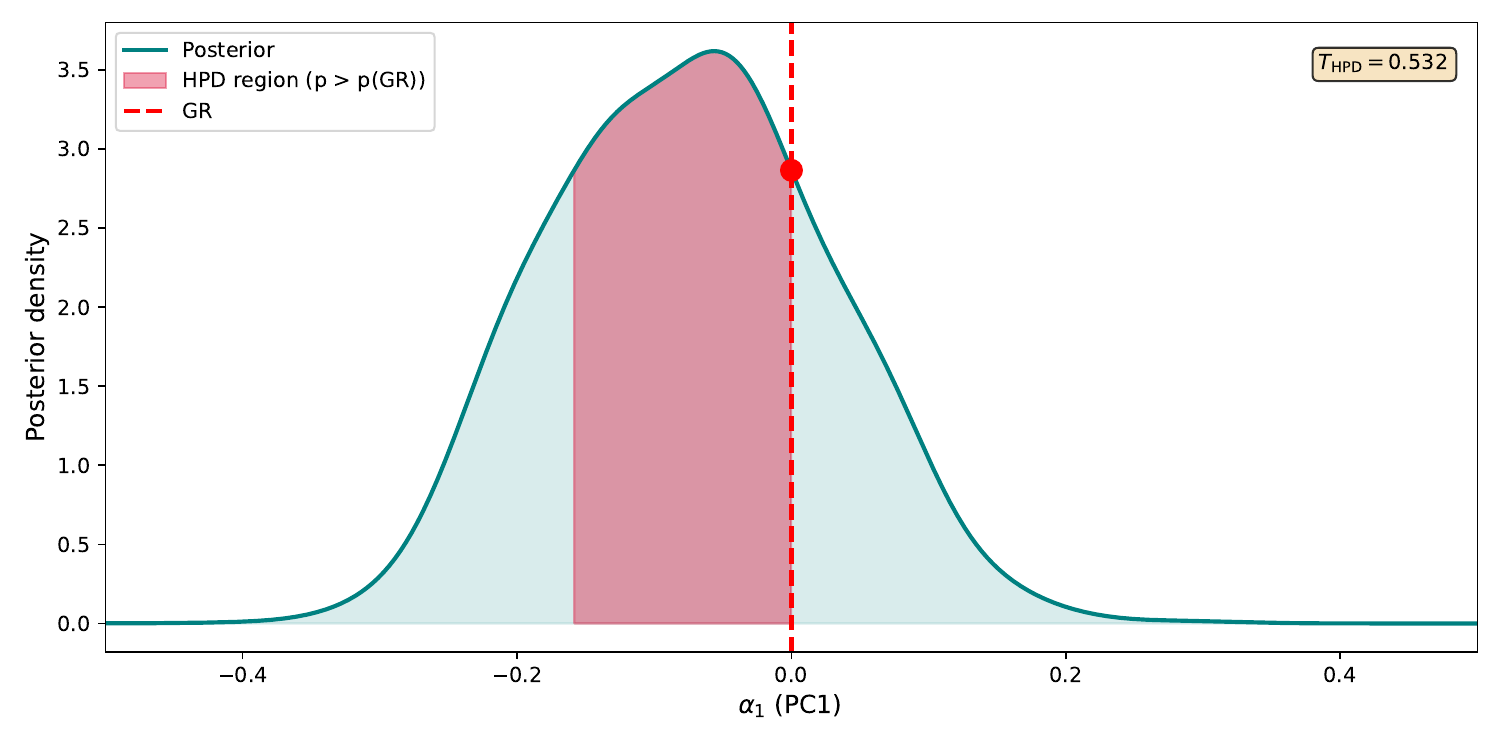}
	\caption{Illustration of the HPD-credibility statistic $T_{\mathrm{HPD}}$ for one GW150914-like GR injection. The teal curve is the PC1 posterior density, and the red dashed line marks the GR value. The shaded region contains points with posterior density greater than that at GR; its integrated probability gives $T_{\mathrm{HPD}}=0.532$.}
	\label{fig:T_HPD_data1}
\end{figure}

The mild offset of the leading PCA mode in GW150914 motivates testing whether comparable offsets can arise from noise fluctuations under GR.
Earlier injection studies in detector noise examined noise-induced PN offsets~\cite{Meidam:2017dgf}.
We therefore perform injection--recovery analyses of GW150914-like GR signals in independent Gaussian-noise realizations. For GW150914 and each GR injection, we construct a separate PCA basis from the corresponding importance-weighted DINGO-IS posterior. We order the modes by increasing posterior variance and evaluate $T_{\mathrm{HPD}}$ for PC1 using the same procedure for the observed and simulated data.
Figure~\ref{fig:T_HPD_data1} illustrates the calculation of $T_{\mathrm{HPD}}$ in Eq.~\eqref{eq:T_HPD} for one injection.
The tighter, GR-consistent posterior of GW250114 provides a useful observational contrast, but does not by itself establish the origin of the GW150914 offset.

\begin{figure*}[!htbp]
	\centering
	\includegraphics[width=\linewidth]{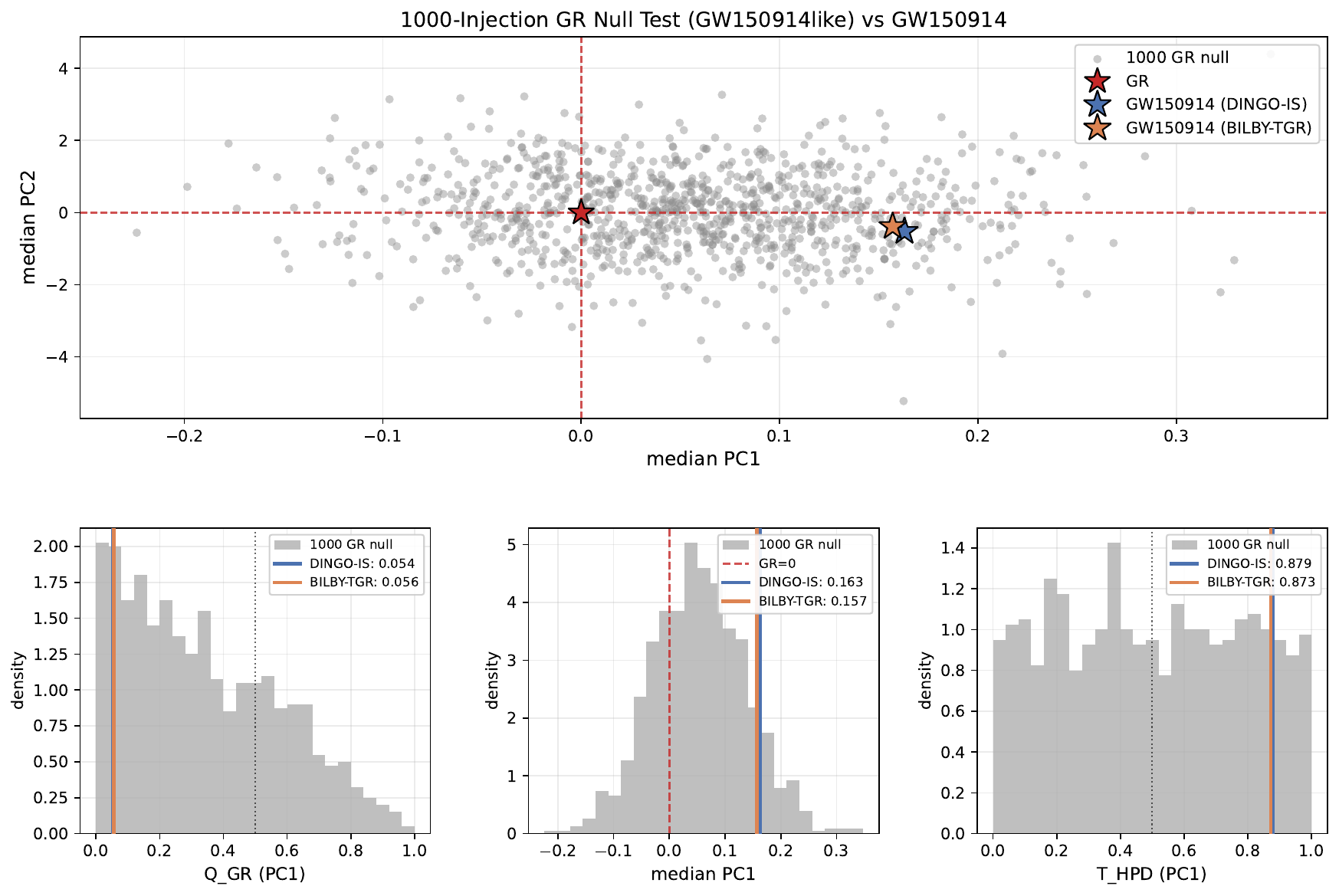}
	\caption{Position of GW150914 within the GR null distributions from 1000 GW150914-like injections with independent Gaussian-noise realizations, all with $\mathrm{ESS}>100$. Each posterior is analyzed in its own PCA basis. PC1 and PC2 denote its two best-constrained modes and need not represent identical PN combinations across data sets. The upper panel shows the posterior medians of the injections (gray), GW150914 from DINGO-IS (blue) and \texttt{bilby-tgr} (orange), and GR (red star). The lower panels show the DINGO-IS null distributions of $\mathcal{Q}_{\mathrm{GR}}$, the PC1 median, and $T_{\mathrm{HPD}}$; vertical lines locate the real-event values. The observed DINGO-IS value $T_{\mathrm{HPD}}^{\mathrm{obs}}=0.879$ gives an empirical upper-tail $p$-value of $11.5\%$.}
	\label{fig:gw150914_in_null1000}
\end{figure*}
We construct a GR null background from GW150914-like injections at a fixed GR-fit configuration: detector-frame chirp mass $M_c=31.3166\,\allowbreak M_\odot$, mass ratio $q=0.9835$, luminosity distance $d_L=425.33$~Mpc, and spin magnitudes $(a_1,\allowbreak a_2)=(0.2483,\allowbreak 0.2641)$. The spin angles are $(\theta_1,\allowbreak \theta_2,\allowbreak \phi_{12},\allowbreak \phi_{JL})=(2.2410,\allowbreak 0.9941,\allowbreak 4.1620,\allowbreak 2.6189)$~rad. We fix the extrinsic parameters to $(\theta_{\rm JN},\allowbreak \phi_{\mathrm{ref}},\allowbreak \alpha,\allowbreak \delta,\allowbreak \psi)=(3.1396,\allowbreak 5.7665,\allowbreak 1.0084,\allowbreak -1.0901,\allowbreak 2.9804)$~rad and set all PN deformation parameters to zero.
We analyze 1000 injections with independent Gaussian-noise realizations, all with $\mathrm{ESS}>100$.
For H1 and L1, we estimate the noise PSDs from randomly selected O1 strain segments and hold them fixed across the injection campaign.
All injections are analyzed with the trained NPE model and the importance-sampling procedure described in Sec.~\ref{subsubsec:GW150914}, using 50,000 21-dimensional proposal samples per injection after synthetic phase sampling.
The selected injections have a median ESS of approximately 894.
The importance-sampling stage uses 64 CPU cores per run, with a median runtime of approximately 10.5 minutes per injection.
Figure~\ref{fig:gw150914_in_null1000} shows the resulting GR null distributions using a separate PCA basis for each data set. The \texttt{bilby-tgr} result is shown for comparison.
Under the adopted GR null hypothesis $H_0$, the upper-tail probability is
\begin{equation}
p=P_{H_0}\!\left(T_{\mathrm{HPD}}\ge T_{\mathrm{HPD}}^{\mathrm{obs}}\right).
\label{eq:gr_null_pvalue}
\end{equation}
Of the 1000 GR injections, 115 reach or exceed the observed $T_{\mathrm{HPD}}$, giving $p=(11.5\pm1.98)\%$ for DINGO-IS. The \texttt{bilby-tgr} result gives $p=(12.2\pm2.03)\%$ against the same null background. The uncertainties are approximate $95\%$ Wald half-widths, $1.96\sqrt{\hat p(1-\hat p)/1000}$, due to the finite number of injections.
As an auxiliary diagnostic, $\mathcal{Q}_{\mathrm{GR}}$ yields an equal-tailed two-sided $p$-value of $21.2\%$, obtained by doubling the smaller empirical tail probability.
Neither test rejects the adopted GR null model at the $5\%$ significance level.

As a robustness check, repeating the test in the fixed basis of zero-noise GR injection No.~9 in Table~\ref{results-16inj} gives an upper-tail $p$-value of $10.0\%$, leaving the conclusion unchanged.
The angles between the injection-specific PC1 eigenvectors and this fixed reference PC1 eigenvector have a median of $0.702^\circ$ ($\cos\theta=0.9999$) and a 95th percentile of $2.029^\circ$ ($\cos\theta=0.9994$).
This narrow angular spread indicates a stable PC1 direction across noise realizations.

Under the adopted Gaussian-noise model and fixed GW150914-like configuration, noise fluctuations can produce offsets comparable to the observed one.
These tests do not determine the origin of the GW150914 offset or quantify additional effects from nonstationary detector noise and waveform-modeling errors.
For example, non-Gaussian noise transients can induce apparent PN deviations~\cite{Kwok:2021zny}.

\section{Conclusion} 
\label{sec:conclusion} 
Our results support the recovery of data-informed combinations of correlated PN deformations through joint inference with DINGO-IS. Using amortized NPE for repeated inference, we also quantify the rarity of the GW150914 offset under the adopted GR noise model.
Using IMRPhenomXPHM waveforms with spin precession and higher-order modes, we jointly inferred GW source parameters and six PN deformation parameters within the TIGER framework, using NPE followed by importance sampling.
After marginalizing over the source parameters, we diagonalized the covariance matrix of the PN deformation posterior to obtain the PCA modes.
We retained modes with a JSD greater than $0.1$ bits between their prior and posterior distributions, restricting the analysis to combinations informed by the data.

We assessed posterior recovery with this framework using zero-noise GR and BGR injections.
The GW150914 model was trained on three million simulations drawn from a parameter range encompassing the event posterior and was also used for the large-scale injection--recovery campaign.
For all 16 GR injections listed in Table~\ref{results-16inj}, the GR value lies within the $90\%$ credible intervals of both the leading and subleading PCA modes, with the nearer, higher-SNR configurations generally yielding tighter constraints.
For the 12 BGR injections listed in Table~\ref{results-12inj}, the posterior medians of the two PCA modes lie close to the projected injected values.
For the uniform positive deformation patterns, the displacement of the leading mode from GR increases with the injected amplitude.
The $0.5$ and increasing patterns exclude GR from the PC1 $90\%$ credible intervals, while all 12 intervals contain the injected values.
The present injection set does not establish sensitivity to resolvable nonzero PC2 deviations.
For all three tested deformation patterns, the 1.5PN term gives the largest absolute contribution when projected onto the GW150914 \texttt{bilby-tgr} PC1 direction.

We applied the same pipeline to the real events GW150914 and GW250114.
For GW150914, the leading PCA mode shows a mild offset from GR, with the GR value near the boundary of its $90\%$ credible interval.
The result agrees with our \texttt{bilby-tgr} analysis and is qualitatively consistent with previous measurements.
The DINGO-IS analysis takes approximately 38 minutes, compared with 177 hours for \texttt{bilby-tgr} with \texttt{dynesty}.
For the higher-SNR event GW250114, we trained a separate model on three million simulations drawn from a narrower prior.
The recovered PCA posteriors are tighter than those of GW150914, consistent with GR, and in agreement with our \texttt{bilby-tgr} analysis.
These GR-consistent results also agree qualitatively with the published LVK conclusion, although the PCA bases differ.
The DINGO-IS analysis takes approximately 4.4 hours, compared with 238 hours for \texttt{bilby-tgr}.

The contrast between the GR-consistent result for GW250114 and the mild offset in GW150914 motivated a noise study.
Using the same GW150914 model, we analyzed 1000 GR injections with independent noise realizations, all with $\mathrm{ESS}>100$. For GW150914 and each injection, we constructed a separate PCA basis from the corresponding DINGO-IS posterior.
Such repeated analyses would be computationally expensive with \texttt{bilby-tgr} and \texttt{dynesty}.
Under the adopted O1-PSD Gaussian-noise model, noise fluctuations produce PCA offsets of varying magnitudes, including offsets larger than that observed in GW150914.
The empirical upper-tail $p$-value for $T_{\mathrm{HPD}}$ is $(11.5\pm1.98)\%$ (approximate $95\%$ Wald interval).
The adopted GR null model is not rejected at the $5\%$ significance level.
Noise fluctuations are therefore a plausible contributor to the observed offset.

The amortized nature of NPE reduces the cost of repeated inference within the training domain.
This approach may also benefit longer signals and more complex sources, such as extreme-mass-ratio inspirals, after suitable extensions.
Potential applications include dedicated studies of individual high-value events involving BGR effects, lensed gravitational waves, or black-hole ringdown quasinormal modes, together with injection-based assessments of their statistical significance.

The present analysis uses a single waveform approximant and restricts the signals to quasicircular binaries.
Waveform-modeling uncertainties may therefore affect the inferred deformation parameters.
Future work should assess these uncertainties using multiple waveform models and extend the framework to eccentric binaries.
Applying the approach to a larger catalog could also support population-level tests of GR.

\section*{Acknowledgments}
T.Y. acknowledges support from the National Natural Science Foundation of China under Grant No. 12575063. H.W. acknowledges support from the National Natural Science Foundation of China under Grant Nos. 12405076 and 12547104. Some of the numerical calculations in this work were performed using the high-performance computing cluster at Wuhan University.

%The calculations in this work used the Python packages \texttt{NumPy}, \texttt{SciPy}, and \texttt{PyTorch}, and the plots were produced with \texttt{Matplotlib}. The analysis also used \texttt{Bilby}, \texttt{bilby-tgr}, \texttt{DINGO-gw}, \texttt{PyCBC}, \texttt{LALSuite}, and \texttt{dynesty}.

\section*{Data Availability}
The datasets generated or analyzed during the current study are available from the corresponding author on reasonable request.

\appendix
\section{Neural-posterior training and calibration}
\label{app:training_settings}

We trained a separate NPE model for each real event. Both models use precessing IMRPhenomXPHM waveforms with six TIGER deformation parameters, an $8\,\mathrm{s}$ data segment, and a frequency range of $20$--$1024\,\mathrm{Hz}$. The SVD representations used for training retain 250 and 300 components for GW150914 and GW250114, respectively. The network architecture largely follows the standard DINGO configuration~\cite{Dax:2021tsq,Dax:2022pxd}; the output includes six PN deformation parameters, and we use the event-specific training strategies described below.

% Declare appendix floats early so wide panels stay near their text.
\begin{figure}[!t]
	\centering
	\includegraphics[width=\linewidth]{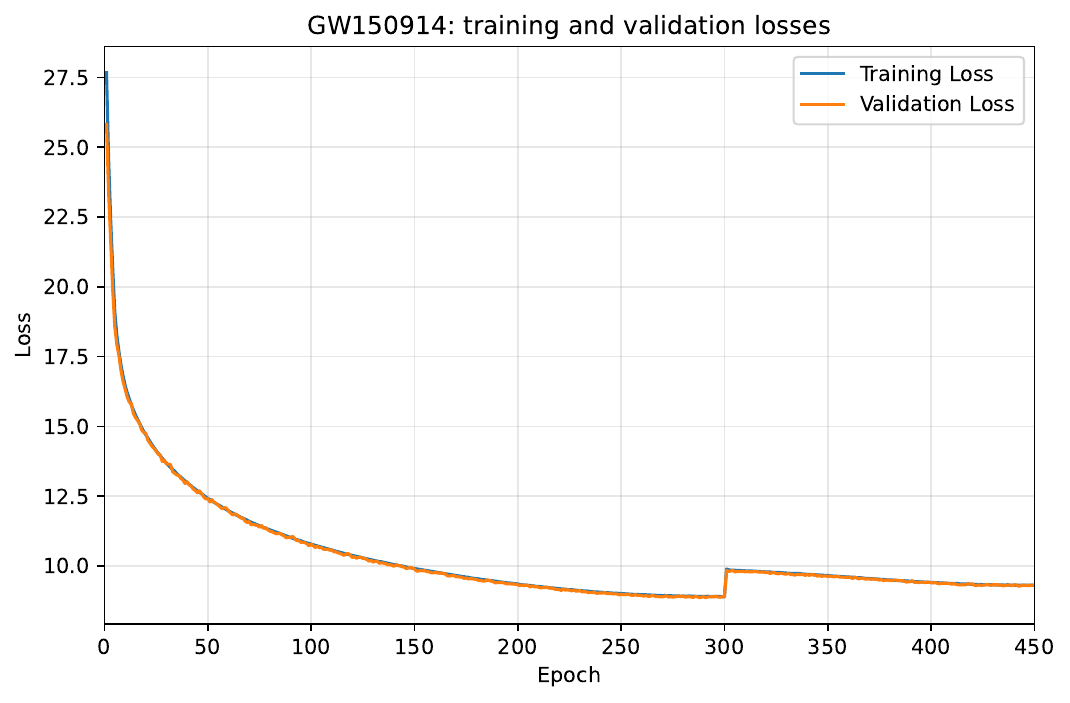}
	\caption{Training and validation losses for the neural posterior estimator used in the GW150914 analysis. The model is trained for 450 epochs on $3\times10^6$ simulated signals with an SVD basis of size 250. The change near epoch 300 marks the transition from training with a fixed fiducial PSD to training with an ensemble of O1 PSDs; the two curves remain closely matched.}
	\label{fig:training_loss}
\end{figure}

\begin{figure}[!t]
	\centering
	\includegraphics[width=\linewidth]{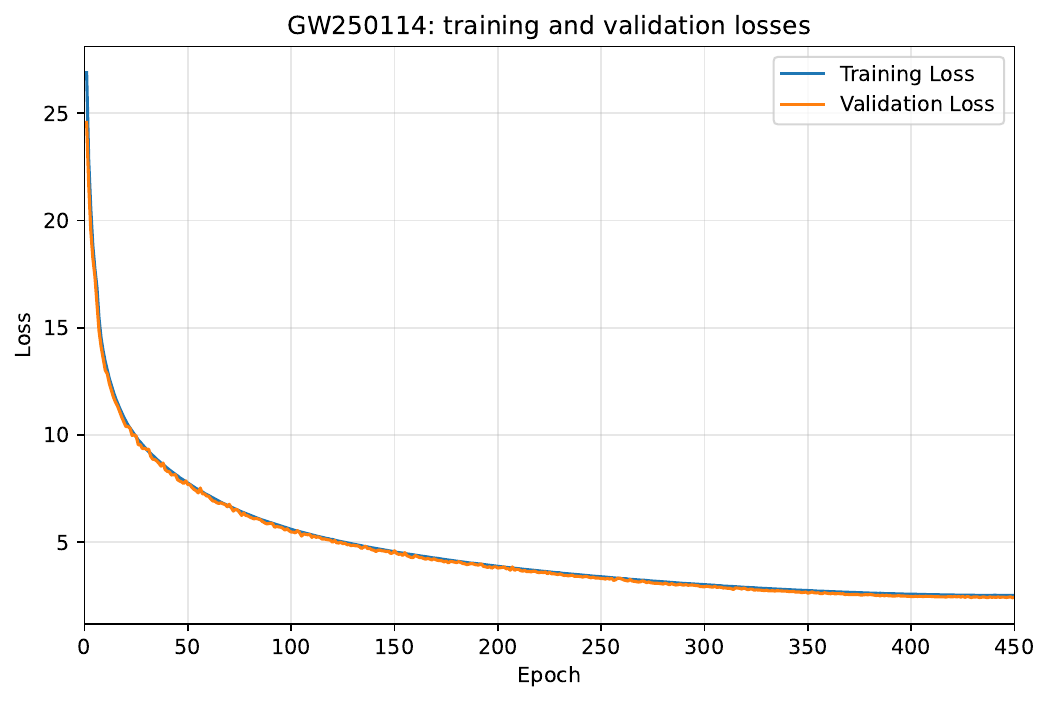}
	\caption{Training and validation losses for the neural posterior estimator used in the GW250114 analysis. The model is trained for 450 epochs on $3\times10^6$ simulated signals drawn from the narrow prior, using a fixed fiducial O4b PSD and an SVD basis of size 300. The training and validation curves remain closely aligned throughout training.}
	\label{fig:training_loss_3e6_model400}
\end{figure}

\begin{figure*}[!t]
	\centering
	\begin{minipage}[t]{0.49\textwidth}
		\centering
		\makebox[\linewidth][l]{\textbf{(a)} GW150914}\par
		\smallskip
		\includegraphics[width=\linewidth]{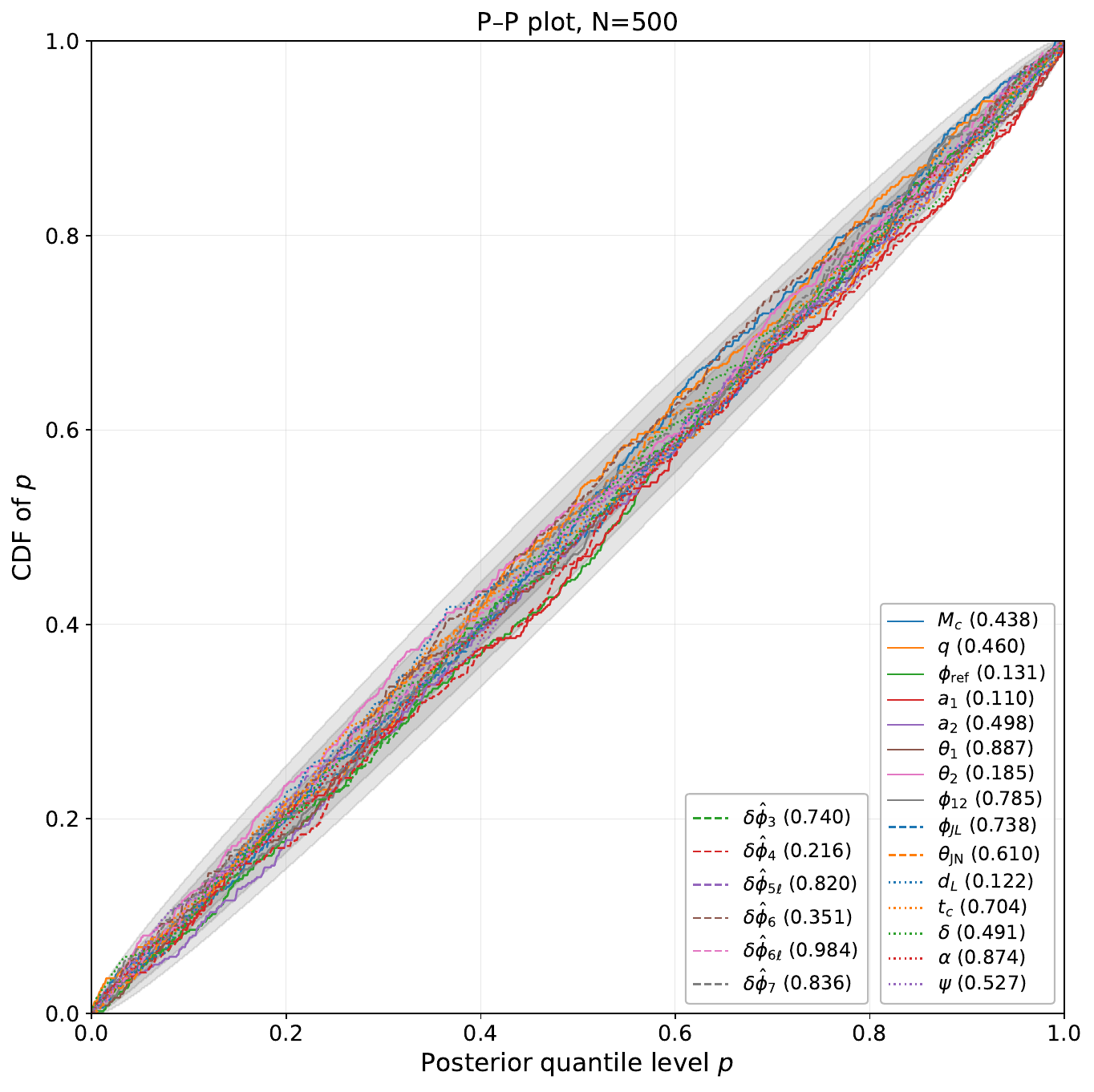}
	\end{minipage}\hfill
	\begin{minipage}[t]{0.49\textwidth}
		\centering
		\makebox[\linewidth][l]{\textbf{(b)} GW250114}\par
		\smallskip
		\includegraphics[width=\linewidth]{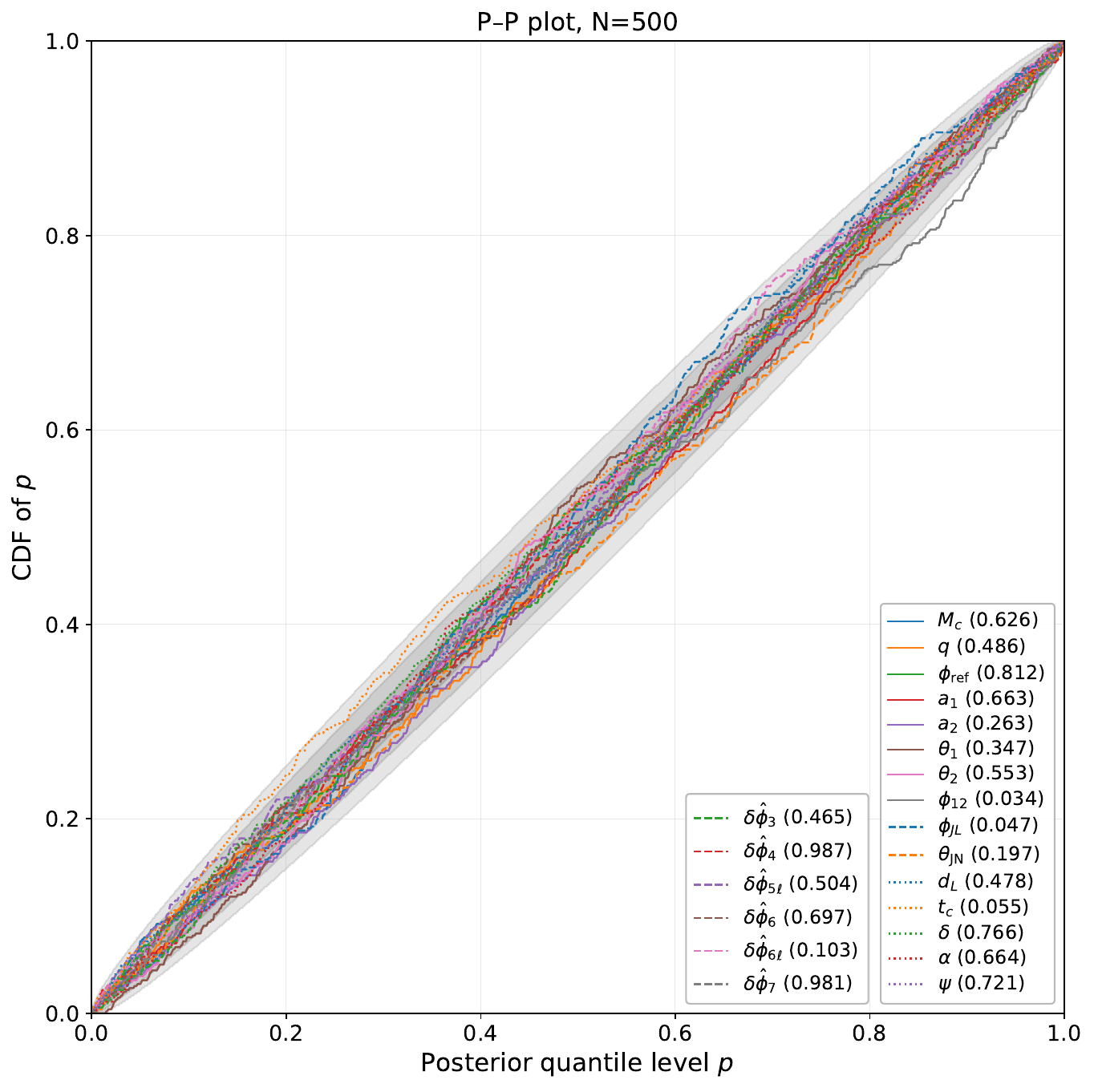}
	\end{minipage}
	\caption{P--P calibration tests of DINGO-IS posteriors for (a) the GW150914 model and (b) the GW250114 model, using 500 simulated signals per model, all with $\mathrm{ESS}>100$. Colored curves show the empirical cumulative distributions of posterior quantiles at the injected values for all 21 parameters; the diagonal denotes ideal calibration, and the shaded bands show the expected sampling fluctuations. Legend entries give single-parameter Kolmogorov--Smirnov $p$-values.}
	\label{fig:pp_calibration}
\end{figure*}

\subsection*{GW150914 model}

For GW150914, we generated $3\times10^6$ waveforms from the broad prior in Table~\ref{tab:GW150914priors}. The model was trained for 450 epochs in two stages. During the first 300 epochs, all examples used a fixed fiducial PSD for each detector. During the remaining 150 epochs, the PSD for each detector was drawn from an ensemble of 400 O1 PSDs. This second stage exposes the network to measured variations in the detector noise. The full 450-epoch training run took $67.5\,\mathrm{h}$ on one RTX 4090 GPU with 96 CPU cores.

Figure~\ref{fig:training_loss} shows that the training and validation losses remain closely matched in both stages. We use the final checkpoint at epoch 450 for the GW150914 inference.

\subsection*{GW250114 model}

For GW250114, we generated $3\times10^6$ waveforms from the narrower prior in Table~\ref{tab:GW250114priors}. These waveforms were used to train a more concentrated proposal for the higher-SNR event. We retain 300 of the 400 stored SVD components for training. The network directly infers 20 parameters, including the six PN deformations, while synthetic phase sampling adds the reference phase during the importance-sampling stage. Training uses one fixed fiducial O4b PSD for each detector throughout all 450 epochs, motivated by preliminary tests suggesting improved importance-sampling efficiency. Training took approximately $67\,\mathrm{h}$ on one RTX 4090 GPU with 32 CPU cores.

The training and validation loss curves in Fig.~\ref{fig:training_loss_3e6_model400} decrease smoothly and remain closely matched. We use the final checkpoint at epoch 450 for the event analysis.

\subsection*{Posterior calibration}

We assess posterior calibration for each model using 500 simulated signals, all with $\mathrm{ESS}>100$. P--P calibration retains the uniform-distance training prior. For each injection, we obtain a 21-dimensional DINGO-IS posterior by importance-weighting 50,000 proposal samples. Figure~\ref{fig:pp_calibration} compares the resulting P--P curves for the GW150914 and GW250114 models.

The P--P curves broadly follow the diagonal and provide marginal calibration diagnostics for the injection sets shown. The real-event analyses reported in Sec.~\ref{sec:results} use the same importance-sampling procedure, with the weights of Eq.~\eqref{eq:is_weights} accounting for differences between the proposal and the target posterior.

\section{PCA bases}
\label{app:pca_details}
\begin{table*}[!t]
\centering
\caption{Coefficients of the two leading event-specific PCA modes for GW150914 and GW250114, constructed from the corresponding real-data \texttt{bilby-tgr} posteriors obtained with \texttt{dynesty}. The columns follow the order of the six PN deformation parameters in Eq.~\eqref{eq:theta_D}; coefficients are rounded to three decimal places.}
\label{tab:pca_coefficients}
\setlength{\tabcolsep}{6pt}
\begin{tabular}{ccrrrrrr}
\toprule
Event & Mode & \multicolumn{1}{c}{$\delta\hat{\phi}_3$} & \multicolumn{1}{c}{$\delta\hat{\phi}_4$} & \multicolumn{1}{c}{$\delta\hat{\phi}_{5\ell}$} & \multicolumn{1}{c}{$\delta\hat{\phi}_6$} & \multicolumn{1}{c}{$\delta\hat{\phi}_{6\ell}$} & \multicolumn{1}{c}{$\delta\hat{\phi}_7$} \\
\midrule
GW150914 & PC1 & $0.904$ & $-0.120$ & $0.337$ & $-0.216$ & $0.050$ & $0.075$ \\
		 & PC2 & $0.407$ & $0.083$ & $-0.583$ & $0.618$ & $-0.107$ & $-0.308$ \\
GW250114 & PC1 & $0.912$ & $-0.120$ & $0.324$ & $-0.206$ & $0.049$ & $0.068$ \\
		 & PC2 & $-0.394$ & $-0.087$ & $0.615$ & $-0.613$ & $0.094$ & $0.274$ \\
\bottomrule
\end{tabular}
\end{table*}
We apply PCA to the six-dimensional PN deformation subspace after marginalizing over the binary parameters. For DINGO-IS posteriors, the mean and covariance are estimated using the normalized weights $\widetilde{w}_i=w_i/\sum_j w_j$. The eigenvalues are ordered from smallest to largest, so PC1 has the smallest variance in the posterior used to construct the basis.

Each eigenvector is defined only up to an overall sign. Following Ref.~\cite{Mahapatra:2025cwk}, we choose the sign such that the component with the largest absolute value is positive. This convention fixes the orientation of the PCA axes without changing their physical interpretation or the conclusions of the GR-consistency tests. We report angles between unit eigenvectors in degrees, using $\theta=\arccos|\boldsymbol{u}\cdot\boldsymbol{v}|$.

For the zero-noise GR and BGR recovery tests, each injection uses its own importance-weighted posterior basis. For each real-event method comparison, both posteriors are projected onto the eigenbasis of the event's \texttt{bilby-tgr} posterior (Sec.~\ref{subsec:real_events}). This common basis need not diagonalize both posterior covariances. For the 1000-injection null test, GW150914 and each injection use their own DINGO-IS posterior basis (Sec.~\ref{subsec:GW150914like_injections}). The fixed basis of zero-noise GR injection No.~9 in Table~\ref{results-16inj} is used only for the robustness check.

The coefficients of the two leading event-specific PCA modes, constructed from each event's real-data \texttt{bilby-tgr} posterior obtained with \texttt{dynesty}, are listed in Table~\ref{tab:pca_coefficients}, with each mode defined by Eq.~\eqref{eq:pca_transformation} and the sign convention given above. For both events, the largest absolute component of PC1 is that of $\delta\hat{\phi}_3$, with a coefficient of approximately 0.91. The largest absolute components of PC2 are those of $\delta\hat{\phi}_{5\ell}$ and $\delta\hat{\phi}_6$. The coefficients show similar leading directions for the two events.

% Switch the URL color back to blue so the arXiv/DOI links in the reference list
% match the citations (the front-matter emails were already typeset in red).
\FloatBarrier
\hypersetup{urlcolor=blue}
\bibliography{references.bib}

%apsrev4-2.bst 2019-01-14 (MD) hand-edited version of apsrev4-1.bst
%Control: key (0)
%Control: author (8) initials jnrlst
%Control: editor formatted (1) identically to author
%Control: production of article title (0) allowed
%Control: page (0) single
%Control: year (1) truncated
%Control: production of eprint (0) enabled
\begin{thebibliography}{61}%
\makeatletter
\providecommand \@ifxundefined [1]{%
 \@ifx{#1\undefined}
}%
\providecommand \@ifnum [1]{%
 \ifnum #1\expandafter \@firstoftwo
 \else \expandafter \@secondoftwo
 \fi
}%
\providecommand \@ifx [1]{%
 \ifx #1\expandafter \@firstoftwo
 \else \expandafter \@secondoftwo
 \fi
}%
\providecommand \natexlab [1]{#1}%
\providecommand \enquote  [1]{``#1''}%
\providecommand \bibnamefont  [1]{#1}%
\providecommand \bibfnamefont [1]{#1}%
\providecommand \citenamefont [1]{#1}%
\providecommand \href@noop [0]{\@secondoftwo}%
\providecommand \href [0]{\begingroup \@sanitize@url \@href}%
\providecommand \@href[1]{\@@startlink{#1}\@@href}%
\providecommand \@@href[1]{\endgroup#1\@@endlink}%
\providecommand \@sanitize@url [0]{\catcode `\\12\catcode `\$12\catcode
  `\&12\catcode `\#12\catcode `\^12\catcode `\_12\catcode `\%12\relax}%
\providecommand \@@startlink[1]{}%
\providecommand \@@endlink[0]{}%
\providecommand \url  [0]{\begingroup\@sanitize@url \@url }%
\providecommand \@url [1]{\endgroup\@href {#1}{\urlprefix }}%
\providecommand \urlprefix  [0]{URL }%
\providecommand \Eprint [0]{\href }%
\providecommand \doibase [0]{https://doi.org/}%
\providecommand \selectlanguage [0]{\@gobble}%
\providecommand \bibinfo  [0]{\@secondoftwo}%
\providecommand \bibfield  [0]{\@secondoftwo}%
\providecommand \translation [1]{[#1]}%
\providecommand \BibitemOpen [0]{}%
\providecommand \bibitemStop [0]{}%
\providecommand \bibitemNoStop [0]{.\EOS\space}%
\providecommand \EOS [0]{\spacefactor3000\relax}%
\providecommand \BibitemShut  [1]{\csname bibitem#1\endcsname}%
\let\auto@bib@innerbib\@empty
%</preamble>
\bibitem [{\citenamefont {Abbott}\ \emph
  {et~al.}(2016{\natexlab{a}})\citenamefont {Abbott} \emph
  {et~al.}}]{LIGOScientific:2016aoc}%
  \BibitemOpen
  \bibfield  {author} {\bibinfo {author} {\bibfnamefont {B.~P.}\ \bibnamefont
  {Abbott}} \emph {et~al.} (\bibinfo {collaboration} {LIGO Scientific,
  Virgo}),\ }\bibfield  {title} {\bibinfo {title} {{Observation of
  Gravitational Waves from a Binary Black Hole Merger}},\ }\href
  {https://doi.org/10.1103/PhysRevLett.116.061102} {\bibfield  {journal}
  {\bibinfo  {journal} {Phys. Rev. Lett.}\ }\textbf {\bibinfo {volume} {116}},\
  \bibinfo {pages} {061102} (\bibinfo {year} {2016}{\natexlab{a}})},\ \Eprint
  {https://arxiv.org/abs/1602.03837} {arXiv:1602.03837 [gr-qc]} \BibitemShut
  {NoStop}%
\bibitem [{\citenamefont {Abbott}\ \emph
  {et~al.}(2016{\natexlab{b}})\citenamefont {Abbott} \emph
  {et~al.}}]{LIGOScientific:2016vlm}%
  \BibitemOpen
  \bibfield  {author} {\bibinfo {author} {\bibfnamefont {B.~P.}\ \bibnamefont
  {Abbott}} \emph {et~al.} (\bibinfo {collaboration} {LIGO Scientific,
  Virgo}),\ }\bibfield  {title} {\bibinfo {title} {{Properties of the Binary
  Black Hole Merger GW150914}},\ }\href
  {https://doi.org/10.1103/PhysRevLett.116.241102} {\bibfield  {journal}
  {\bibinfo  {journal} {Phys. Rev. Lett.}\ }\textbf {\bibinfo {volume} {116}},\
  \bibinfo {pages} {241102} (\bibinfo {year} {2016}{\natexlab{b}})},\ \Eprint
  {https://arxiv.org/abs/1602.03840} {arXiv:1602.03840 [gr-qc]} \BibitemShut
  {NoStop}%
\bibitem [{\citenamefont {Abbott}\ \emph
  {et~al.}(2016{\natexlab{c}})\citenamefont {Abbott} \emph
  {et~al.}}]{LIGOScientific:2016lio}%
  \BibitemOpen
  \bibfield  {author} {\bibinfo {author} {\bibfnamefont {B.~P.}\ \bibnamefont
  {Abbott}} \emph {et~al.} (\bibinfo {collaboration} {LIGO Scientific,
  Virgo}),\ }\bibfield  {title} {\bibinfo {title} {{Tests of general relativity
  with GW150914}},\ }\href {https://doi.org/10.1103/PhysRevLett.116.221101}
  {\bibfield  {journal} {\bibinfo  {journal} {Phys. Rev. Lett.}\ }\textbf
  {\bibinfo {volume} {116}},\ \bibinfo {pages} {221101} (\bibinfo {year}
  {2016}{\natexlab{c}})},\ \bibinfo {note} {[Erratum: Phys.Rev.Lett. 121,
  129902 (2018)]},\ \Eprint {https://arxiv.org/abs/1602.03841}
  {arXiv:1602.03841 [gr-qc]} \BibitemShut {NoStop}%
\bibitem [{\citenamefont {Blanchet}(2006)}]{Blanchet:2006jqj}%
  \BibitemOpen
  \bibfield  {author} {\bibinfo {author} {\bibfnamefont {L.}~\bibnamefont
  {Blanchet}},\ }\bibfield  {title} {\bibinfo {title} {{Gravitational Radiation
  from Post-Newtonian Sources and Inspiralling Compact Binaries}},\ }\href
  {https://doi.org/10.12942/lrr-2006-4} {\bibfield  {journal} {\bibinfo
  {journal} {Living Rev. Rel.}\ }\textbf {\bibinfo {volume} {9}},\ \bibinfo
  {pages} {4} (\bibinfo {year} {2006})}\BibitemShut {NoStop}%
\bibitem [{\citenamefont {Blanchet}\ \emph {et~al.}(1995)\citenamefont
  {Blanchet}, \citenamefont {Damour},\ and\ \citenamefont
  {Iyer}}]{Blanchet:1995fg}%
  \BibitemOpen
  \bibfield  {author} {\bibinfo {author} {\bibfnamefont {L.}~\bibnamefont
  {Blanchet}}, \bibinfo {author} {\bibfnamefont {T.}~\bibnamefont {Damour}},\
  and\ \bibinfo {author} {\bibfnamefont {B.~R.}\ \bibnamefont {Iyer}},\
  }\bibfield  {title} {\bibinfo {title} {{Gravitational waves from inspiralling
  compact binaries: Energy loss and wave form to second postNewtonian order}},\
  }\href {https://doi.org/10.1103/PhysRevD.51.5360} {\bibfield  {journal}
  {\bibinfo  {journal} {Phys. Rev. D}\ }\textbf {\bibinfo {volume} {51}},\
  \bibinfo {pages} {5360} (\bibinfo {year} {1995})},\ \bibinfo {note}
  {[Erratum: Phys.Rev.D 54, 1860 (1996)]},\ \Eprint
  {https://arxiv.org/abs/gr-qc/9501029} {arXiv:gr-qc/9501029} \BibitemShut
  {NoStop}%
\bibitem [{\citenamefont {Tahura}\ \emph {et~al.}(2019)\citenamefont {Tahura},
  \citenamefont {Yagi},\ and\ \citenamefont {Carson}}]{Tahura:2019dgr}%
  \BibitemOpen
  \bibfield  {author} {\bibinfo {author} {\bibfnamefont {S.}~\bibnamefont
  {Tahura}}, \bibinfo {author} {\bibfnamefont {K.}~\bibnamefont {Yagi}},\ and\
  \bibinfo {author} {\bibfnamefont {Z.}~\bibnamefont {Carson}},\ }\bibfield
  {title} {\bibinfo {title} {{Testing Gravity with Gravitational Waves from
  Binary Black Hole Mergers: Contributions from Amplitude Corrections}},\
  }\href {https://doi.org/10.1103/PhysRevD.100.104001} {\bibfield  {journal}
  {\bibinfo  {journal} {Phys. Rev. D}\ }\textbf {\bibinfo {volume} {100}},\
  \bibinfo {pages} {104001} (\bibinfo {year} {2019})},\ \Eprint
  {https://arxiv.org/abs/1907.10059} {arXiv:1907.10059 [gr-qc]} \BibitemShut
  {NoStop}%
\bibitem [{\citenamefont {Mishra}\ \emph {et~al.}(2010)\citenamefont {Mishra},
  \citenamefont {Arun}, \citenamefont {Iyer},\ and\ \citenamefont
  {Sathyaprakash}}]{Mishra:2010tp}%
  \BibitemOpen
  \bibfield  {author} {\bibinfo {author} {\bibfnamefont {C.~K.}\ \bibnamefont
  {Mishra}}, \bibinfo {author} {\bibfnamefont {K.~G.}\ \bibnamefont {Arun}},
  \bibinfo {author} {\bibfnamefont {B.~R.}\ \bibnamefont {Iyer}},\ and\
  \bibinfo {author} {\bibfnamefont {B.~S.}\ \bibnamefont {Sathyaprakash}},\
  }\bibfield  {title} {\bibinfo {title} {{Parametrized tests of post-Newtonian
  theory using Advanced LIGO and Einstein Telescope}},\ }\href
  {https://doi.org/10.1103/PhysRevD.82.064010} {\bibfield  {journal} {\bibinfo
  {journal} {Phys. Rev. D}\ }\textbf {\bibinfo {volume} {82}},\ \bibinfo
  {pages} {064010} (\bibinfo {year} {2010})},\ \Eprint
  {https://arxiv.org/abs/1005.0304} {arXiv:1005.0304 [gr-qc]} \BibitemShut
  {NoStop}%
\bibitem [{\citenamefont {Agathos}\ \emph {et~al.}(2014)\citenamefont
  {Agathos}, \citenamefont {Del~Pozzo}, \citenamefont {Li}, \citenamefont {Van
  Den~Broeck}, \citenamefont {Veitch},\ and\ \citenamefont
  {Vitale}}]{Agathos:2013upa}%
  \BibitemOpen
  \bibfield  {author} {\bibinfo {author} {\bibfnamefont {M.}~\bibnamefont
  {Agathos}}, \bibinfo {author} {\bibfnamefont {W.}~\bibnamefont {Del~Pozzo}},
  \bibinfo {author} {\bibfnamefont {T.~G.~F.}\ \bibnamefont {Li}}, \bibinfo
  {author} {\bibfnamefont {C.}~\bibnamefont {Van Den~Broeck}}, \bibinfo
  {author} {\bibfnamefont {J.}~\bibnamefont {Veitch}},\ and\ \bibinfo {author}
  {\bibfnamefont {S.}~\bibnamefont {Vitale}},\ }\bibfield  {title} {\bibinfo
  {title} {{TIGER: A data analysis pipeline for testing the strong-field
  dynamics of general relativity with gravitational wave signals from
  coalescing compact binaries}},\ }\href
  {https://doi.org/10.1103/PhysRevD.89.082001} {\bibfield  {journal} {\bibinfo
  {journal} {Phys. Rev. D}\ }\textbf {\bibinfo {volume} {89}},\ \bibinfo
  {pages} {082001} (\bibinfo {year} {2014})},\ \Eprint
  {https://arxiv.org/abs/1311.0420} {arXiv:1311.0420 [gr-qc]} \BibitemShut
  {NoStop}%
\bibitem [{\citenamefont {Mehta}\ \emph {et~al.}(2023)\citenamefont {Mehta},
  \citenamefont {Buonanno}, \citenamefont {Cotesta}, \citenamefont {Ghosh},
  \citenamefont {Sennett},\ and\ \citenamefont {Steinhoff}}]{Mehta:2022pcn}%
  \BibitemOpen
  \bibfield  {author} {\bibinfo {author} {\bibfnamefont {A.~K.}\ \bibnamefont
  {Mehta}}, \bibinfo {author} {\bibfnamefont {A.}~\bibnamefont {Buonanno}},
  \bibinfo {author} {\bibfnamefont {R.}~\bibnamefont {Cotesta}}, \bibinfo
  {author} {\bibfnamefont {A.}~\bibnamefont {Ghosh}}, \bibinfo {author}
  {\bibfnamefont {N.}~\bibnamefont {Sennett}},\ and\ \bibinfo {author}
  {\bibfnamefont {J.}~\bibnamefont {Steinhoff}},\ }\bibfield  {title} {\bibinfo
  {title} {{Tests of general relativity with gravitational-wave observations
  using a flexible theory-independent method}},\ }\href
  {https://doi.org/10.1103/PhysRevD.107.044020} {\bibfield  {journal} {\bibinfo
   {journal} {Phys. Rev. D}\ }\textbf {\bibinfo {volume} {107}},\ \bibinfo
  {pages} {044020} (\bibinfo {year} {2023})},\ \Eprint
  {https://arxiv.org/abs/2203.13937} {arXiv:2203.13937 [gr-qc]} \BibitemShut
  {NoStop}%
\bibitem [{\citenamefont {Blanchet}\ \emph {et~al.}(2023)\citenamefont
  {Blanchet}, \citenamefont {Faye}, \citenamefont {Henry}, \citenamefont
  {Larrouturou},\ and\ \citenamefont {Trestini}}]{Blanchet:2023bwj}%
  \BibitemOpen
  \bibfield  {author} {\bibinfo {author} {\bibfnamefont {L.}~\bibnamefont
  {Blanchet}}, \bibinfo {author} {\bibfnamefont {G.}~\bibnamefont {Faye}},
  \bibinfo {author} {\bibfnamefont {Q.}~\bibnamefont {Henry}}, \bibinfo
  {author} {\bibfnamefont {F.}~\bibnamefont {Larrouturou}},\ and\ \bibinfo
  {author} {\bibfnamefont {D.}~\bibnamefont {Trestini}},\ }\bibfield  {title}
  {\bibinfo {title} {{Gravitational-Wave Phasing of Quasicircular Compact
  Binary Systems to the Fourth-and-a-Half Post-Newtonian Order}},\ }\href
  {https://doi.org/10.1103/PhysRevLett.131.121402} {\bibfield  {journal}
  {\bibinfo  {journal} {Phys. Rev. Lett.}\ }\textbf {\bibinfo {volume} {131}},\
  \bibinfo {pages} {121402} (\bibinfo {year} {2023})},\ \Eprint
  {https://arxiv.org/abs/2304.11185} {arXiv:2304.11185 [gr-qc]} \BibitemShut
  {NoStop}%
\bibitem [{\citenamefont {Meidam}\ \emph {et~al.}(2018)\citenamefont {Meidam}
  \emph {et~al.}}]{Meidam:2017dgf}%
  \BibitemOpen
  \bibfield  {author} {\bibinfo {author} {\bibfnamefont {J.}~\bibnamefont
  {Meidam}} \emph {et~al.},\ }\bibfield  {title} {\bibinfo {title}
  {{Parametrized tests of the strong-field dynamics of general relativity using
  gravitational wave signals from coalescing binary black holes: Fast
  likelihood calculations and sensitivity of the method}},\ }\href
  {https://doi.org/10.1103/PhysRevD.97.044033} {\bibfield  {journal} {\bibinfo
  {journal} {Phys. Rev. D}\ }\textbf {\bibinfo {volume} {97}},\ \bibinfo
  {pages} {044033} (\bibinfo {year} {2018})},\ \Eprint
  {https://arxiv.org/abs/1712.08772} {arXiv:1712.08772 [gr-qc]} \BibitemShut
  {NoStop}%
\bibitem [{\citenamefont {Pratten}\ \emph {et~al.}(2021)\citenamefont {Pratten}
  \emph {et~al.}}]{Pratten:2020ceb}%
  \BibitemOpen
  \bibfield  {author} {\bibinfo {author} {\bibfnamefont {G.}~\bibnamefont
  {Pratten}} \emph {et~al.},\ }\bibfield  {title} {\bibinfo {title}
  {{Computationally efficient models for the dominant and subdominant harmonic
  modes of precessing binary black holes}},\ }\href
  {https://doi.org/10.1103/PhysRevD.103.104056} {\bibfield  {journal} {\bibinfo
   {journal} {Phys. Rev. D}\ }\textbf {\bibinfo {volume} {103}},\ \bibinfo
  {pages} {104056} (\bibinfo {year} {2021})},\ \Eprint
  {https://arxiv.org/abs/2004.06503} {arXiv:2004.06503 [gr-qc]} \BibitemShut
  {NoStop}%
\bibitem [{\citenamefont {Ashton}\ \emph {et~al.}(2019)\citenamefont {Ashton}
  \emph {et~al.}}]{Ashton:2018jfp}%
  \BibitemOpen
  \bibfield  {author} {\bibinfo {author} {\bibfnamefont {G.}~\bibnamefont
  {Ashton}} \emph {et~al.},\ }\bibfield  {title} {\bibinfo {title} {{BILBY: A
  user-friendly Bayesian inference library for gravitational-wave astronomy}},\
  }\href {https://doi.org/10.3847/1538-4365/ab06fc} {\bibfield  {journal}
  {\bibinfo  {journal} {Astrophys. J. Suppl.}\ }\textbf {\bibinfo {volume}
  {241}},\ \bibinfo {pages} {27} (\bibinfo {year} {2019})},\ \Eprint
  {https://arxiv.org/abs/1811.02042} {arXiv:1811.02042 [astro-ph.IM]}
  \BibitemShut {NoStop}%
\bibitem [{\citenamefont {Speagle}(2020)}]{Speagle:2019ivv}%
  \BibitemOpen
  \bibfield  {author} {\bibinfo {author} {\bibfnamefont {J.~S.}\ \bibnamefont
  {Speagle}},\ }\bibfield  {title} {\bibinfo {title} {{dynesty: a dynamic
  nested sampling package for estimating Bayesian posteriors and evidences}},\
  }\href {https://doi.org/10.1093/mnras/staa278} {\bibfield  {journal}
  {\bibinfo  {journal} {Mon. Not. Roy. Astron. Soc.}\ }\textbf {\bibinfo
  {volume} {493}},\ \bibinfo {pages} {3132} (\bibinfo {year} {2020})},\ \Eprint
  {https://arxiv.org/abs/1904.02180} {arXiv:1904.02180 [astro-ph.IM]}
  \BibitemShut {NoStop}%
\bibitem [{\citenamefont {Cranmer}\ \emph {et~al.}(2020)\citenamefont
  {Cranmer}, \citenamefont {Brehmer},\ and\ \citenamefont
  {Louppe}}]{Cranmer:2019eaq}%
  \BibitemOpen
  \bibfield  {author} {\bibinfo {author} {\bibfnamefont {K.}~\bibnamefont
  {Cranmer}}, \bibinfo {author} {\bibfnamefont {J.}~\bibnamefont {Brehmer}},\
  and\ \bibinfo {author} {\bibfnamefont {G.}~\bibnamefont {Louppe}},\
  }\bibfield  {title} {\bibinfo {title} {{The frontier of simulation-based
  inference}},\ }\href {https://doi.org/10.1073/pnas.1912789117} {\bibfield
  {journal} {\bibinfo  {journal} {Proc. Nat. Acad. Sci.}\ }\textbf {\bibinfo
  {volume} {117}},\ \bibinfo {pages} {30055} (\bibinfo {year} {2020})},\
  \Eprint {https://arxiv.org/abs/1911.01429} {arXiv:1911.01429 [stat.ML]}
  \BibitemShut {NoStop}%
\bibitem [{\citenamefont {Liang}\ and\ \citenamefont
  {Wang}(2026)}]{Liang:2025yjw}%
  \BibitemOpen
  \bibfield  {author} {\bibinfo {author} {\bibfnamefont {B.}~\bibnamefont
  {Liang}}\ and\ \bibinfo {author} {\bibfnamefont {H.}~\bibnamefont {Wang}},\
  }\bibfield  {title} {\bibinfo {title} {{Recent advances in simulation-based
  inference for gravitational wave data analysis}},\ }\href
  {https://doi.org/10.61977/ati2025020} {\bibfield  {journal} {\bibinfo
  {journal} {Astronomical Techniques and Instruments}\ }\textbf {\bibinfo
  {volume} {3}},\ \bibinfo {pages} {93} (\bibinfo {year} {2026})},\ \Eprint
  {https://arxiv.org/abs/2507.11192} {arXiv:2507.11192 [gr-qc]} \BibitemShut
  {NoStop}%
\bibitem [{\citenamefont {Chua}\ and\ \citenamefont
  {Vallisneri}(2020)}]{Chua:2019wwt}%
  \BibitemOpen
  \bibfield  {author} {\bibinfo {author} {\bibfnamefont {A.~J.~K.}\
  \bibnamefont {Chua}}\ and\ \bibinfo {author} {\bibfnamefont {M.}~\bibnamefont
  {Vallisneri}},\ }\bibfield  {title} {\bibinfo {title} {{Learning Bayesian
  posteriors with neural networks for gravitational-wave inference}},\ }\href
  {https://doi.org/10.1103/PhysRevLett.124.041102} {\bibfield  {journal}
  {\bibinfo  {journal} {Phys. Rev. Lett.}\ }\textbf {\bibinfo {volume} {124}},\
  \bibinfo {pages} {041102} (\bibinfo {year} {2020})},\ \Eprint
  {https://arxiv.org/abs/1909.05966} {arXiv:1909.05966 [gr-qc]} \BibitemShut
  {NoStop}%
\bibitem [{\citenamefont {Green}\ \emph {et~al.}(2020)\citenamefont {Green},
  \citenamefont {Simpson},\ and\ \citenamefont {Gair}}]{Green:2020hst}%
  \BibitemOpen
  \bibfield  {author} {\bibinfo {author} {\bibfnamefont {S.~R.}\ \bibnamefont
  {Green}}, \bibinfo {author} {\bibfnamefont {C.}~\bibnamefont {Simpson}},\
  and\ \bibinfo {author} {\bibfnamefont {J.}~\bibnamefont {Gair}},\ }\bibfield
  {title} {\bibinfo {title} {{Gravitational-wave parameter estimation with
  autoregressive neural network flows}},\ }\href
  {https://doi.org/10.1103/PhysRevD.102.104057} {\bibfield  {journal} {\bibinfo
   {journal} {Phys. Rev. D}\ }\textbf {\bibinfo {volume} {102}},\ \bibinfo
  {pages} {104057} (\bibinfo {year} {2020})},\ \Eprint
  {https://arxiv.org/abs/2002.07656} {arXiv:2002.07656 [astro-ph.IM]}
  \BibitemShut {NoStop}%
\bibitem [{\citenamefont {Dax}\ \emph {et~al.}(2021{\natexlab{a}})\citenamefont
  {Dax}, \citenamefont {Green}, \citenamefont {Gair}, \citenamefont {Macke},
  \citenamefont {Buonanno},\ and\ \citenamefont {Sch{\"o}lkopf}}]{Dax:2021tsq}%
  \BibitemOpen
  \bibfield  {author} {\bibinfo {author} {\bibfnamefont {M.}~\bibnamefont
  {Dax}}, \bibinfo {author} {\bibfnamefont {S.~R.}\ \bibnamefont {Green}},
  \bibinfo {author} {\bibfnamefont {J.}~\bibnamefont {Gair}}, \bibinfo {author}
  {\bibfnamefont {J.~H.}\ \bibnamefont {Macke}}, \bibinfo {author}
  {\bibfnamefont {A.}~\bibnamefont {Buonanno}},\ and\ \bibinfo {author}
  {\bibfnamefont {B.}~\bibnamefont {Sch{\"o}lkopf}},\ }\bibfield  {title}
  {\bibinfo {title} {{Real-Time Gravitational Wave Science with Neural
  Posterior Estimation}},\ }\href
  {https://doi.org/10.1103/PhysRevLett.127.241103} {\bibfield  {journal}
  {\bibinfo  {journal} {Phys. Rev. Lett.}\ }\textbf {\bibinfo {volume} {127}},\
  \bibinfo {pages} {241103} (\bibinfo {year} {2021}{\natexlab{a}})},\ \Eprint
  {https://arxiv.org/abs/2106.12594} {arXiv:2106.12594 [gr-qc]} \BibitemShut
  {NoStop}%
\bibitem [{\citenamefont {Dax}\ \emph {et~al.}(2023)\citenamefont {Dax},
  \citenamefont {Green}, \citenamefont {Gair}, \citenamefont {P{\"u}rrer},
  \citenamefont {Wildberger}, \citenamefont {Macke}, \citenamefont {Buonanno},\
  and\ \citenamefont {Sch{\"o}lkopf}}]{Dax:2022pxd}%
  \BibitemOpen
  \bibfield  {author} {\bibinfo {author} {\bibfnamefont {M.}~\bibnamefont
  {Dax}}, \bibinfo {author} {\bibfnamefont {S.~R.}\ \bibnamefont {Green}},
  \bibinfo {author} {\bibfnamefont {J.}~\bibnamefont {Gair}}, \bibinfo {author}
  {\bibfnamefont {M.}~\bibnamefont {P{\"u}rrer}}, \bibinfo {author}
  {\bibfnamefont {J.}~\bibnamefont {Wildberger}}, \bibinfo {author}
  {\bibfnamefont {J.~H.}\ \bibnamefont {Macke}}, \bibinfo {author}
  {\bibfnamefont {A.}~\bibnamefont {Buonanno}},\ and\ \bibinfo {author}
  {\bibfnamefont {B.}~\bibnamefont {Sch{\"o}lkopf}},\ }\bibfield  {title}
  {\bibinfo {title} {{Neural Importance Sampling for Rapid and Reliable
  Gravitational-Wave Inference}},\ }\href
  {https://doi.org/10.1103/PhysRevLett.130.171403} {\bibfield  {journal}
  {\bibinfo  {journal} {Phys. Rev. Lett.}\ }\textbf {\bibinfo {volume} {130}},\
  \bibinfo {pages} {171403} (\bibinfo {year} {2023})},\ \Eprint
  {https://arxiv.org/abs/2210.05686} {arXiv:2210.05686 [gr-qc]} \BibitemShut
  {NoStop}%
\bibitem [{\citenamefont {Kofler}\ \emph {et~al.}(2026)\citenamefont {Kofler},
  \citenamefont {Dax}, \citenamefont {Green}, \citenamefont {Wildberger},
  \citenamefont {Gupte}, \citenamefont {Macke}, \citenamefont {Gair},
  \citenamefont {Buonanno},\ and\ \citenamefont
  {Sch{\"o}lkopf}}]{Kofler:2025dux}%
  \BibitemOpen
  \bibfield  {author} {\bibinfo {author} {\bibfnamefont {A.}~\bibnamefont
  {Kofler}}, \bibinfo {author} {\bibfnamefont {M.}~\bibnamefont {Dax}},
  \bibinfo {author} {\bibfnamefont {S.~R.}\ \bibnamefont {Green}}, \bibinfo
  {author} {\bibfnamefont {J.}~\bibnamefont {Wildberger}}, \bibinfo {author}
  {\bibfnamefont {N.}~\bibnamefont {Gupte}}, \bibinfo {author} {\bibfnamefont
  {J.~H.}\ \bibnamefont {Macke}}, \bibinfo {author} {\bibfnamefont
  {J.}~\bibnamefont {Gair}}, \bibinfo {author} {\bibfnamefont {A.}~\bibnamefont
  {Buonanno}},\ and\ \bibinfo {author} {\bibfnamefont {B.}~\bibnamefont
  {Sch{\"o}lkopf}},\ }\bibfield  {title} {\bibinfo {title} {{Flexible
  Gravitational-Wave Parameter Estimation with Transformers}},\ }\href
  {https://doi.org/10.1103/n3ld-b4wv} {\bibfield  {journal} {\bibinfo
  {journal} {Phys. Rev. Lett.}\ }\textbf {\bibinfo {volume} {136}},\ \bibinfo
  {pages} {251401} (\bibinfo {year} {2026})},\ \Eprint
  {https://arxiv.org/abs/2512.02968} {arXiv:2512.02968 [gr-qc]} \BibitemShut
  {NoStop}%
\bibitem [{\citenamefont {Xie}\ \emph {et~al.}(2024)\citenamefont {Xie},
  \citenamefont {Chatterjee}, \citenamefont {Narayan},\ and\ \citenamefont
  {Yunes}}]{Xie:2024ubm}%
  \BibitemOpen
  \bibfield  {author} {\bibinfo {author} {\bibfnamefont {Y.}~\bibnamefont
  {Xie}}, \bibinfo {author} {\bibfnamefont {D.}~\bibnamefont {Chatterjee}},
  \bibinfo {author} {\bibfnamefont {G.}~\bibnamefont {Narayan}},\ and\ \bibinfo
  {author} {\bibfnamefont {N.}~\bibnamefont {Yunes}},\ }\bibfield  {title}
  {\bibinfo {title} {{Neural post-Einsteinian framework for efficient
  theory-agnostic tests of general relativity with gravitational waves}},\
  }\href {https://doi.org/10.1103/PhysRevD.110.024036} {\bibfield  {journal}
  {\bibinfo  {journal} {Phys. Rev. D}\ }\textbf {\bibinfo {volume} {110}},\
  \bibinfo {pages} {024036} (\bibinfo {year} {2024})},\ \Eprint
  {https://arxiv.org/abs/2403.18936} {arXiv:2403.18936 [gr-qc]} \BibitemShut
  {NoStop}%
\bibitem [{\citenamefont {Xie}\ \emph {et~al.}(2026)\citenamefont {Xie},
  \citenamefont {Narayan},\ and\ \citenamefont {Yunes}}]{Xie:2025voe}%
  \BibitemOpen
  \bibfield  {author} {\bibinfo {author} {\bibfnamefont {Y.}~\bibnamefont
  {Xie}}, \bibinfo {author} {\bibfnamefont {G.}~\bibnamefont {Narayan}},\ and\
  \bibinfo {author} {\bibfnamefont {N.}~\bibnamefont {Yunes}},\ }\bibfield
  {title} {\bibinfo {title} {{Neural Post-Einsteinian Test of General
  Relativity with the Third Gravitational Wave Transient Catalog}},\ }\href
  {https://doi.org/10.1103/n1pb-jt1n} {\bibfield  {journal} {\bibinfo
  {journal} {Phys. Rev. Lett.}\ }\textbf {\bibinfo {volume} {136}},\ \bibinfo
  {pages} {181402} (\bibinfo {year} {2026})},\ \Eprint
  {https://arxiv.org/abs/2510.02515} {arXiv:2510.02515 [gr-qc]} \BibitemShut
  {NoStop}%
\bibitem [{\citenamefont {Zhang}\ \emph {et~al.}(2026)\citenamefont {Zhang},
  \citenamefont {Sun}, \citenamefont {Xiong}, \citenamefont {Liu},
  \citenamefont {Wang}, \citenamefont {Jin}, \citenamefont {Zhang},\ and\
  \citenamefont {Zhang}}]{Zhang:2026okw}%
  \BibitemOpen
  \bibfield  {author} {\bibinfo {author} {\bibfnamefont {Y.-X.}\ \bibnamefont
  {Zhang}}, \bibinfo {author} {\bibfnamefont {T.-Y.}\ \bibnamefont {Sun}},
  \bibinfo {author} {\bibfnamefont {C.-Y.}\ \bibnamefont {Xiong}}, \bibinfo
  {author} {\bibfnamefont {S.-T.}\ \bibnamefont {Liu}}, \bibinfo {author}
  {\bibfnamefont {Y.-X.}\ \bibnamefont {Wang}}, \bibinfo {author}
  {\bibfnamefont {S.-J.}\ \bibnamefont {Jin}}, \bibinfo {author} {\bibfnamefont
  {J.-F.}\ \bibnamefont {Zhang}},\ and\ \bibinfo {author} {\bibfnamefont
  {X.}~\bibnamefont {Zhang}},\ }\bibfield  {title} {\bibinfo {title}
  {{Factorized neural posterior estimation for rapid and reliable inference of
  parameterized post-Einsteinian deviation parameters in gravitational
  waves}},\ }\href {https://doi.org/10.1088/1572-9494/ae8f1e} {\bibfield
  {journal} {\bibinfo  {journal} {Commun. Theor. Phys.}\ }\textbf {\bibinfo
  {volume} {78}},\ \bibinfo {pages} {115401} (\bibinfo {year} {2026})},\
  \Eprint {https://arxiv.org/abs/2602.01319} {arXiv:2602.01319 [astro-ph.IM]}
  \BibitemShut {NoStop}%
\bibitem [{\citenamefont {Wang}\ \emph {et~al.}(2025)\citenamefont {Wang},
  \citenamefont {Wei}, \citenamefont {Li}, \citenamefont {Sun}, \citenamefont
  {Jin}, \citenamefont {Wang}, \citenamefont {Cui}, \citenamefont {Zhang},\
  and\ \citenamefont {Zhang}}]{Wang:2024oei}%
  \BibitemOpen
  \bibfield  {author} {\bibinfo {author} {\bibfnamefont {Y.-X.}\ \bibnamefont
  {Wang}}, \bibinfo {author} {\bibfnamefont {X.}~\bibnamefont {Wei}}, \bibinfo
  {author} {\bibfnamefont {C.-Y.}\ \bibnamefont {Li}}, \bibinfo {author}
  {\bibfnamefont {T.-Y.}\ \bibnamefont {Sun}}, \bibinfo {author} {\bibfnamefont
  {S.-J.}\ \bibnamefont {Jin}}, \bibinfo {author} {\bibfnamefont
  {H.}~\bibnamefont {Wang}}, \bibinfo {author} {\bibfnamefont {J.-L.}\
  \bibnamefont {Cui}}, \bibinfo {author} {\bibfnamefont {J.-F.}\ \bibnamefont
  {Zhang}},\ and\ \bibinfo {author} {\bibfnamefont {X.}~\bibnamefont {Zhang}},\
  }\bibfield  {title} {\bibinfo {title} {{Search for exotic gravitational wave
  signals beyond general relativity using deep learning}},\ }\href
  {https://doi.org/10.1103/f85k-wtph} {\bibfield  {journal} {\bibinfo
  {journal} {Phys. Rev. D}\ }\textbf {\bibinfo {volume} {112}},\ \bibinfo
  {pages} {024030} (\bibinfo {year} {2025})},\ \Eprint
  {https://arxiv.org/abs/2410.20129} {arXiv:2410.20129 [gr-qc]} \BibitemShut
  {NoStop}%
\bibitem [{\citenamefont {Crisostomi}\ \emph {et~al.}(2023)\citenamefont
  {Crisostomi}, \citenamefont {Dey}, \citenamefont {Barausse},\ and\
  \citenamefont {Trotta}}]{Crisostomi:2023tle}%
  \BibitemOpen
  \bibfield  {author} {\bibinfo {author} {\bibfnamefont {M.}~\bibnamefont
  {Crisostomi}}, \bibinfo {author} {\bibfnamefont {K.}~\bibnamefont {Dey}},
  \bibinfo {author} {\bibfnamefont {E.}~\bibnamefont {Barausse}},\ and\
  \bibinfo {author} {\bibfnamefont {R.}~\bibnamefont {Trotta}},\ }\bibfield
  {title} {\bibinfo {title} {{Neural posterior estimation with guaranteed exact
  coverage: The ringdown of GW150914}},\ }\href
  {https://doi.org/10.1103/PhysRevD.108.044029} {\bibfield  {journal} {\bibinfo
   {journal} {Phys. Rev. D}\ }\textbf {\bibinfo {volume} {108}},\ \bibinfo
  {pages} {044029} (\bibinfo {year} {2023})},\ \Eprint
  {https://arxiv.org/abs/2305.18528} {arXiv:2305.18528 [gr-qc]} \BibitemShut
  {NoStop}%
\bibitem [{\citenamefont {Srinivasan}\ \emph {et~al.}(2024)\citenamefont
  {Srinivasan}, \citenamefont {Crisostomi}, \citenamefont {Trotta},
  \citenamefont {Barausse},\ and\ \citenamefont
  {Breschi}}]{Srinivasan:2024uax}%
  \BibitemOpen
  \bibfield  {author} {\bibinfo {author} {\bibfnamefont {R.}~\bibnamefont
  {Srinivasan}}, \bibinfo {author} {\bibfnamefont {M.}~\bibnamefont
  {Crisostomi}}, \bibinfo {author} {\bibfnamefont {R.}~\bibnamefont {Trotta}},
  \bibinfo {author} {\bibfnamefont {E.}~\bibnamefont {Barausse}},\ and\
  \bibinfo {author} {\bibfnamefont {M.}~\bibnamefont {Breschi}},\ }\bibfield
  {title} {\bibinfo {title} {{Bayesian evidence estimation from posterior
  samples with normalizing flows}},\ }\href
  {https://doi.org/10.1103/PhysRevD.110.123007} {\bibfield  {journal} {\bibinfo
   {journal} {Phys. Rev. D}\ }\textbf {\bibinfo {volume} {110}},\ \bibinfo
  {pages} {123007} (\bibinfo {year} {2024})},\ \Eprint
  {https://arxiv.org/abs/2404.12294} {arXiv:2404.12294 [stat.ML]} \BibitemShut
  {NoStop}%
\bibitem [{\citenamefont {Pacilio}\ \emph {et~al.}(2024)\citenamefont
  {Pacilio}, \citenamefont {Bhagwat},\ and\ \citenamefont
  {Cotesta}}]{Pacilio:2024qcq}%
  \BibitemOpen
  \bibfield  {author} {\bibinfo {author} {\bibfnamefont {C.}~\bibnamefont
  {Pacilio}}, \bibinfo {author} {\bibfnamefont {S.}~\bibnamefont {Bhagwat}},\
  and\ \bibinfo {author} {\bibfnamefont {R.}~\bibnamefont {Cotesta}},\
  }\bibfield  {title} {\bibinfo {title} {{Simulation-based inference of black
  hole ringdowns in the time domain}},\ }\href
  {https://doi.org/10.1103/PhysRevD.110.083010} {\bibfield  {journal} {\bibinfo
   {journal} {Phys. Rev. D}\ }\textbf {\bibinfo {volume} {110}},\ \bibinfo
  {pages} {083010} (\bibinfo {year} {2024})},\ \Eprint
  {https://arxiv.org/abs/2404.11373} {arXiv:2404.11373 [gr-qc]} \BibitemShut
  {NoStop}%
\bibitem [{\citenamefont {Chan}\ \emph
  {et~al.}(2026{\natexlab{a}})\citenamefont {Chan}, \citenamefont
  {Maga{\~n}a~Zertuche}, \citenamefont {Ezquiaga}, \citenamefont {Lo},
  \citenamefont {Vujeva},\ and\ \citenamefont {Bowman}}]{Chan:2025pdf}%
  \BibitemOpen
  \bibfield  {author} {\bibinfo {author} {\bibfnamefont {J.~C.~L.}\
  \bibnamefont {Chan}}, \bibinfo {author} {\bibfnamefont {L.}~\bibnamefont
  {Maga{\~n}a~Zertuche}}, \bibinfo {author} {\bibfnamefont {J.~M.}\
  \bibnamefont {Ezquiaga}}, \bibinfo {author} {\bibfnamefont {R.~K.~L.}\
  \bibnamefont {Lo}}, \bibinfo {author} {\bibfnamefont {L.}~\bibnamefont
  {Vujeva}},\ and\ \bibinfo {author} {\bibfnamefont {J.}~\bibnamefont
  {Bowman}},\ }\bibfield  {title} {\bibinfo {title} {{Identification and
  characterization of distorted gravitational waves by lensing using deep
  learning}},\ }\href {https://doi.org/10.1103/8cz1-kl6n} {\bibfield  {journal}
  {\bibinfo  {journal} {Phys. Rev. D}\ }\textbf {\bibinfo {volume} {113}},\
  \bibinfo {pages} {024041} (\bibinfo {year} {2026}{\natexlab{a}})},\ \Eprint
  {https://arxiv.org/abs/2511.07186} {arXiv:2511.07186 [gr-qc]} \BibitemShut
  {NoStop}%
\bibitem [{\citenamefont {Chan}\ \emph
  {et~al.}(2026{\natexlab{b}})\citenamefont {Chan}, \citenamefont {Ezquiaga},
  \citenamefont {Lo}, \citenamefont {Bowman}, \citenamefont
  {Maga{\~n}a~Zertuche},\ and\ \citenamefont {Vujeva}}]{Chan:2025kyu}%
  \BibitemOpen
  \bibfield  {author} {\bibinfo {author} {\bibfnamefont {J.~C.~L.}\
  \bibnamefont {Chan}}, \bibinfo {author} {\bibfnamefont {J.~M.}\ \bibnamefont
  {Ezquiaga}}, \bibinfo {author} {\bibfnamefont {R.~K.~L.}\ \bibnamefont {Lo}},
  \bibinfo {author} {\bibfnamefont {J.}~\bibnamefont {Bowman}}, \bibinfo
  {author} {\bibfnamefont {L.}~\bibnamefont {Maga{\~n}a~Zertuche}},\ and\
  \bibinfo {author} {\bibfnamefont {L.}~\bibnamefont {Vujeva}},\ }\bibfield
  {title} {\bibinfo {title} {{Discovering gravitational waveform distortions
  from lensing: A deep dive into GW231123}},\ }\href
  {https://doi.org/10.1103/5292-948q} {\bibfield  {journal} {\bibinfo
  {journal} {Phys. Rev. D}\ }\textbf {\bibinfo {volume} {114}},\ \bibinfo
  {pages} {L021502} (\bibinfo {year} {2026}{\natexlab{b}})},\ \Eprint
  {https://arxiv.org/abs/2512.16916} {arXiv:2512.16916 [gr-qc]} \BibitemShut
  {NoStop}%
\bibitem [{\citenamefont {Stachurski}\ \emph {et~al.}(2024)\citenamefont
  {Stachurski}, \citenamefont {Messenger},\ and\ \citenamefont
  {Hendry}}]{Stachurski:2023ntw}%
  \BibitemOpen
  \bibfield  {author} {\bibinfo {author} {\bibfnamefont {F.}~\bibnamefont
  {Stachurski}}, \bibinfo {author} {\bibfnamefont {C.}~\bibnamefont
  {Messenger}},\ and\ \bibinfo {author} {\bibfnamefont {M.}~\bibnamefont
  {Hendry}},\ }\bibfield  {title} {\bibinfo {title} {{Cosmological inference
  using gravitational waves and normalizing flows}},\ }\href
  {https://doi.org/10.1103/PhysRevD.109.123547} {\bibfield  {journal} {\bibinfo
   {journal} {Phys. Rev. D}\ }\textbf {\bibinfo {volume} {109}},\ \bibinfo
  {pages} {123547} (\bibinfo {year} {2024})},\ \Eprint
  {https://arxiv.org/abs/2310.13405} {arXiv:2310.13405 [gr-qc]} \BibitemShut
  {NoStop}%
\bibitem [{\citenamefont {Leyde}\ \emph {et~al.}(2024)\citenamefont {Leyde},
  \citenamefont {Green}, \citenamefont {Toubiana},\ and\ \citenamefont
  {Gair}}]{Leyde:2023iof}%
  \BibitemOpen
  \bibfield  {author} {\bibinfo {author} {\bibfnamefont {K.}~\bibnamefont
  {Leyde}}, \bibinfo {author} {\bibfnamefont {S.~R.}\ \bibnamefont {Green}},
  \bibinfo {author} {\bibfnamefont {A.}~\bibnamefont {Toubiana}},\ and\
  \bibinfo {author} {\bibfnamefont {J.}~\bibnamefont {Gair}},\ }\bibfield
  {title} {\bibinfo {title} {{Gravitational wave populations and cosmology with
  neural posterior estimation}},\ }\href
  {https://doi.org/10.1103/PhysRevD.109.064056} {\bibfield  {journal} {\bibinfo
   {journal} {Phys. Rev. D}\ }\textbf {\bibinfo {volume} {109}},\ \bibinfo
  {pages} {064056} (\bibinfo {year} {2024})},\ \Eprint
  {https://arxiv.org/abs/2311.12093} {arXiv:2311.12093 [gr-qc]} \BibitemShut
  {NoStop}%
\bibitem [{\citenamefont {Xiong}\ \emph {et~al.}(2025)\citenamefont {Xiong},
  \citenamefont {Sun}, \citenamefont {Zhang},\ and\ \citenamefont
  {Zhang}}]{Xiong:2024gpx}%
  \BibitemOpen
  \bibfield  {author} {\bibinfo {author} {\bibfnamefont {C.-Y.}\ \bibnamefont
  {Xiong}}, \bibinfo {author} {\bibfnamefont {T.-Y.}\ \bibnamefont {Sun}},
  \bibinfo {author} {\bibfnamefont {J.-F.}\ \bibnamefont {Zhang}},\ and\
  \bibinfo {author} {\bibfnamefont {X.}~\bibnamefont {Zhang}},\ }\bibfield
  {title} {\bibinfo {title} {{Robust inference of gravitational wave source
  parameters in the presence of noise transients using normalizing flows}},\
  }\href {https://doi.org/10.1103/PhysRevD.111.024019} {\bibfield  {journal}
  {\bibinfo  {journal} {Phys. Rev. D}\ }\textbf {\bibinfo {volume} {111}},\
  \bibinfo {pages} {024019} (\bibinfo {year} {2025})},\ \Eprint
  {https://arxiv.org/abs/2405.09475} {arXiv:2405.09475 [gr-qc]} \BibitemShut
  {NoStop}%
\bibitem [{\citenamefont {Kobyzev}\ \emph {et~al.}(2021)\citenamefont
  {Kobyzev}, \citenamefont {Prince},\ and\ \citenamefont
  {Brubaker}}]{Kobyzev:2019ydm}%
  \BibitemOpen
  \bibfield  {author} {\bibinfo {author} {\bibfnamefont {I.}~\bibnamefont
  {Kobyzev}}, \bibinfo {author} {\bibfnamefont {S.~J.~D.}\ \bibnamefont
  {Prince}},\ and\ \bibinfo {author} {\bibfnamefont {M.~A.}\ \bibnamefont
  {Brubaker}},\ }\bibfield  {title} {\bibinfo {title} {{Normalizing Flows: An
  Introduction and Review of Current Methods}},\ }\href
  {https://doi.org/10.1109/tpami.2020.2992934} {\bibfield  {journal} {\bibinfo
  {journal} {IEEE Trans. Pattern Anal. Machine Intell.}\ }\textbf {\bibinfo
  {volume} {43}},\ \bibinfo {pages} {3964} (\bibinfo {year} {2021})},\ \Eprint
  {https://arxiv.org/abs/1908.09257} {arXiv:1908.09257 [stat.ML]} \BibitemShut
  {NoStop}%
\bibitem [{\citenamefont {Papamakarios}\ \emph {et~al.}(2021)\citenamefont
  {Papamakarios}, \citenamefont {Nalisnick}, \citenamefont {Rezende},
  \citenamefont {Mohamed},\ and\ \citenamefont
  {Lakshminarayanan}}]{Papamakarios:2019fms}%
  \BibitemOpen
  \bibfield  {author} {\bibinfo {author} {\bibfnamefont {G.}~\bibnamefont
  {Papamakarios}}, \bibinfo {author} {\bibfnamefont {E.}~\bibnamefont
  {Nalisnick}}, \bibinfo {author} {\bibfnamefont {D.~J.}\ \bibnamefont
  {Rezende}}, \bibinfo {author} {\bibfnamefont {S.}~\bibnamefont {Mohamed}},\
  and\ \bibinfo {author} {\bibfnamefont {B.}~\bibnamefont {Lakshminarayanan}},\
  }\bibfield  {title} {\bibinfo {title} {{Normalizing Flows for Probabilistic
  Modeling and Inference}},\ }\href
  {https://www.jmlr.org/papers/v22/19-1028.html} {\bibfield  {journal}
  {\bibinfo  {journal} {J. Machine Learning Res.}\ }\textbf {\bibinfo {volume}
  {22}},\ \bibinfo {pages} {1} (\bibinfo {year} {2021})},\ \Eprint
  {https://arxiv.org/abs/1912.02762} {arXiv:1912.02762 [stat.ML]} \BibitemShut
  {NoStop}%
\bibitem [{\citenamefont {Pinchbeck}\ \emph {et~al.}(2026)\citenamefont
  {Pinchbeck}, \citenamefont {Thrane}, \citenamefont {Balazs},\ and\
  \citenamefont {Lasky}}]{Pinchbeck:2026hup}%
  \BibitemOpen
  \bibfield  {author} {\bibinfo {author} {\bibfnamefont {L.}~\bibnamefont
  {Pinchbeck}}, \bibinfo {author} {\bibfnamefont {E.}~\bibnamefont {Thrane}},
  \bibinfo {author} {\bibfnamefont {C.}~\bibnamefont {Balazs}},\ and\ \bibinfo
  {author} {\bibfnamefont {P.}~\bibnamefont {Lasky}},\ }\href@noop {} {\bibinfo
  {title} {{$\texttt{BilbyFlow}$: user-friendly neural posterior estimation for
  gravitational-wave astronomy}}} (\bibinfo {year} {2026}),\ \Eprint
  {https://arxiv.org/abs/2609.00766} {arXiv:2609.00766 [astro-ph.IM]}
  \BibitemShut {NoStop}%
\bibitem [{\citenamefont {Dax}\ \emph {et~al.}(2021{\natexlab{b}})\citenamefont
  {Dax}, \citenamefont {Wildberger}, \citenamefont {Buchholz}, \citenamefont
  {Green}, \citenamefont {Macke},\ and\ \citenamefont
  {Sch{\"o}lkopf}}]{dingo_gw_repo}%
  \BibitemOpen
  \bibfield  {author} {\bibinfo {author} {\bibfnamefont {M.}~\bibnamefont
  {Dax}}, \bibinfo {author} {\bibfnamefont {J.}~\bibnamefont {Wildberger}},
  \bibinfo {author} {\bibfnamefont {S.}~\bibnamefont {Buchholz}}, \bibinfo
  {author} {\bibfnamefont {S.~R.}\ \bibnamefont {Green}}, \bibinfo {author}
  {\bibfnamefont {J.~H.}\ \bibnamefont {Macke}},\ and\ \bibinfo {author}
  {\bibfnamefont {B.}~\bibnamefont {Sch{\"o}lkopf}},\ }\href@noop {} {\bibinfo
  {title} {{dingo‑gw}: Neural posterior estimation for gravitational‑wave
  inference}},\ \bibinfo {howpublished}
  {\url{https://github.com/dingo-gw/dingo}} (\bibinfo {year}
  {2021}{\natexlab{b}}),\ \bibinfo {note} {gitHub repository}\BibitemShut
  {NoStop}%
\bibitem [{\citenamefont {Abbott}\ \emph
  {et~al.}(2019{\natexlab{a}})\citenamefont {Abbott} \emph
  {et~al.}}]{LIGOScientific:2018dkp}%
  \BibitemOpen
  \bibfield  {author} {\bibinfo {author} {\bibfnamefont {B.~P.}\ \bibnamefont
  {Abbott}} \emph {et~al.} (\bibinfo {collaboration} {LIGO Scientific,
  Virgo}),\ }\bibfield  {title} {\bibinfo {title} {{Tests of General Relativity
  with GW170817}},\ }\href {https://doi.org/10.1103/PhysRevLett.123.011102}
  {\bibfield  {journal} {\bibinfo  {journal} {Phys. Rev. Lett.}\ }\textbf
  {\bibinfo {volume} {123}},\ \bibinfo {pages} {011102} (\bibinfo {year}
  {2019}{\natexlab{a}})},\ \Eprint {https://arxiv.org/abs/1811.00364}
  {arXiv:1811.00364 [gr-qc]} \BibitemShut {NoStop}%
\bibitem [{\citenamefont {Abbott}\ \emph
  {et~al.}(2019{\natexlab{b}})\citenamefont {Abbott} \emph
  {et~al.}}]{LIGOScientific:2019fpa}%
  \BibitemOpen
  \bibfield  {author} {\bibinfo {author} {\bibfnamefont {B.~P.}\ \bibnamefont
  {Abbott}} \emph {et~al.} (\bibinfo {collaboration} {LIGO Scientific,
  Virgo}),\ }\bibfield  {title} {\bibinfo {title} {{Tests of General Relativity
  with the Binary Black Hole Signals from the LIGO-Virgo Catalog GWTC-1}},\
  }\href {https://doi.org/10.1103/PhysRevD.100.104036} {\bibfield  {journal}
  {\bibinfo  {journal} {Phys. Rev. D}\ }\textbf {\bibinfo {volume} {100}},\
  \bibinfo {pages} {104036} (\bibinfo {year} {2019}{\natexlab{b}})},\ \Eprint
  {https://arxiv.org/abs/1903.04467} {arXiv:1903.04467 [gr-qc]} \BibitemShut
  {NoStop}%
\bibitem [{\citenamefont {Abbott}\ \emph {et~al.}(2025)\citenamefont {Abbott}
  \emph {et~al.}}]{LIGOScientific:2021sio}%
  \BibitemOpen
  \bibfield  {author} {\bibinfo {author} {\bibfnamefont {R.}~\bibnamefont
  {Abbott}} \emph {et~al.} (\bibinfo {collaboration} {LIGO Scientific, VIRGO,
  KAGRA}),\ }\bibfield  {title} {\bibinfo {title} {{Tests of General Relativity
  with GWTC-3}},\ }\href {https://doi.org/10.1103/PhysRevD.112.084080}
  {\bibfield  {journal} {\bibinfo  {journal} {Phys. Rev. D}\ }\textbf {\bibinfo
  {volume} {112}},\ \bibinfo {pages} {084080} (\bibinfo {year} {2025})},\
  \Eprint {https://arxiv.org/abs/2112.06861} {arXiv:2112.06861 [gr-qc]}
  \BibitemShut {NoStop}%
\bibitem [{\citenamefont {Mahapatra}\ \emph {et~al.}(2025)\citenamefont
  {Mahapatra} \emph {et~al.}}]{Mahapatra:2025cwk}%
  \BibitemOpen
  \bibfield  {author} {\bibinfo {author} {\bibfnamefont {P.}~\bibnamefont
  {Mahapatra}} \emph {et~al.},\ }\bibfield  {title} {\bibinfo {title}
  {{Confronting general relativity with principal component analysis:
  Simulations and results from GWTC-3 events}},\ }\href
  {https://doi.org/10.1103/c1sj-jc4v} {\bibfield  {journal} {\bibinfo
  {journal} {Phys. Rev. D}\ }\textbf {\bibinfo {volume} {112}},\ \bibinfo
  {pages} {104007} (\bibinfo {year} {2025})},\ \Eprint
  {https://arxiv.org/abs/2508.06862} {arXiv:2508.06862 [gr-qc]} \BibitemShut
  {NoStop}%
\bibitem [{\citenamefont {Heisenberg}\ and\ \citenamefont
  {Hemmatyar}(2026)}]{Heisenberg:2026Identifiability}%
  \BibitemOpen
  \bibfield  {author} {\bibinfo {author} {\bibfnamefont {L.}~\bibnamefont
  {Heisenberg}}\ and\ \bibinfo {author} {\bibfnamefont {S.}~\bibnamefont
  {Hemmatyar}},\ }\href@noop {} {\bibinfo {title} {{Identifiability Limits of
  Gravitational Wave Phase Deviations: Multiclass Classification with a
  Multihead Neural Network}}} (\bibinfo {year} {2026}),\ \Eprint
  {https://arxiv.org/abs/2609.31993} {arXiv:2609.31993 [gr-qc]} \BibitemShut
  {NoStop}%
\bibitem [{\citenamefont {Pai}\ and\ \citenamefont {Arun}(2013)}]{Pai:2012mv}%
  \BibitemOpen
  \bibfield  {author} {\bibinfo {author} {\bibfnamefont {A.}~\bibnamefont
  {Pai}}\ and\ \bibinfo {author} {\bibfnamefont {K.~G.}\ \bibnamefont {Arun}},\
  }\bibfield  {title} {\bibinfo {title} {{Singular value decomposition in
  parametrised tests of post-Newtonian theory}},\ }\href
  {https://doi.org/10.1088/0264-9381/30/2/025011} {\bibfield  {journal}
  {\bibinfo  {journal} {Class. Quant. Grav.}\ }\textbf {\bibinfo {volume}
  {30}},\ \bibinfo {pages} {025011} (\bibinfo {year} {2013})},\ \Eprint
  {https://arxiv.org/abs/1207.1943} {arXiv:1207.1943 [gr-qc]} \BibitemShut
  {NoStop}%
\bibitem [{\citenamefont {Seymour}\ \emph {et~al.}(2026)\citenamefont
  {Seymour}, \citenamefont {Golomb},\ and\ \citenamefont
  {Chen}}]{Seymour:2026bjg}%
  \BibitemOpen
  \bibfield  {author} {\bibinfo {author} {\bibfnamefont {B.~C.}\ \bibnamefont
  {Seymour}}, \bibinfo {author} {\bibfnamefont {J.}~\bibnamefont {Golomb}},\
  and\ \bibinfo {author} {\bibfnamefont {Y.}~\bibnamefont {Chen}},\ }\bibfield
  {title} {\bibinfo {title} {{Inspiral tests of general relativity and waveform
  geometry}},\ }\href {https://doi.org/10.1103/nw3l-qld4} {\bibfield  {journal}
  {\bibinfo  {journal} {Phys. Rev. D}\ }\textbf {\bibinfo {volume} {114}},\
  \bibinfo {pages} {024059} (\bibinfo {year} {2026})},\ \Eprint
  {https://arxiv.org/abs/2602.17524} {arXiv:2602.17524 [gr-qc]} \BibitemShut
  {NoStop}%
\bibitem [{\citenamefont {Niu}\ \emph {et~al.}(2024)\citenamefont {Niu},
  \citenamefont {Ma}, \citenamefont {Chen}, \citenamefont {Feng},\ and\
  \citenamefont {Zhao}}]{Niu:2023kzz}%
  \BibitemOpen
  \bibfield  {author} {\bibinfo {author} {\bibfnamefont {R.}~\bibnamefont
  {Niu}}, \bibinfo {author} {\bibfnamefont {Z.-C.}\ \bibnamefont {Ma}},
  \bibinfo {author} {\bibfnamefont {J.-M.}\ \bibnamefont {Chen}}, \bibinfo
  {author} {\bibfnamefont {C.}~\bibnamefont {Feng}},\ and\ \bibinfo {author}
  {\bibfnamefont {W.}~\bibnamefont {Zhao}},\ }\bibfield  {title} {\bibinfo
  {title} {{Multi-parameter tests of general relativity using Bayesian
  parameter estimation with principal component analysis for LISA}},\ }\href
  {https://doi.org/10.1016/j.rinp.2024.107407} {\bibfield  {journal} {\bibinfo
  {journal} {Results Phys.}\ }\textbf {\bibinfo {volume} {57}},\ \bibinfo
  {pages} {107407} (\bibinfo {year} {2024})},\ \Eprint
  {https://arxiv.org/abs/2310.14195} {arXiv:2310.14195 [gr-qc]} \BibitemShut
  {NoStop}%
\bibitem [{\citenamefont {Saleem}\ \emph {et~al.}(2022)\citenamefont {Saleem},
  \citenamefont {Datta}, \citenamefont {Arun},\ and\ \citenamefont
  {Sathyaprakash}}]{Saleem:2021nsb}%
  \BibitemOpen
  \bibfield  {author} {\bibinfo {author} {\bibfnamefont {M.}~\bibnamefont
  {Saleem}}, \bibinfo {author} {\bibfnamefont {S.}~\bibnamefont {Datta}},
  \bibinfo {author} {\bibfnamefont {K.~G.}\ \bibnamefont {Arun}},\ and\
  \bibinfo {author} {\bibfnamefont {B.~S.}\ \bibnamefont {Sathyaprakash}},\
  }\bibfield  {title} {\bibinfo {title} {{Parametrized tests of post-Newtonian
  theory using principal component analysis}},\ }\href
  {https://doi.org/10.1103/PhysRevD.105.084062} {\bibfield  {journal} {\bibinfo
   {journal} {Phys. Rev. D}\ }\textbf {\bibinfo {volume} {105}},\ \bibinfo
  {pages} {084062} (\bibinfo {year} {2022})},\ \Eprint
  {https://arxiv.org/abs/2110.10147} {arXiv:2110.10147 [gr-qc]} \BibitemShut
  {NoStop}%
\bibitem [{\citenamefont {Abac}\ \emph
  {et~al.}(2026{\natexlab{a}})\citenamefont {Abac} \emph
  {et~al.}}]{LIGOScientific:2026fcf}%
  \BibitemOpen
  \bibfield  {author} {\bibinfo {author} {\bibfnamefont {A.~G.}\ \bibnamefont
  {Abac}} \emph {et~al.} (\bibinfo {collaboration} {LIGO Scientific, VIRGO,
  KAGRA}),\ }\href@noop {} {\bibinfo {title} {{GWTC-4.0: Tests of General
  Relativity. II. Parameterized Tests}}} (\bibinfo {year}
  {2026}{\natexlab{a}}),\ \Eprint {https://arxiv.org/abs/2603.19020}
  {arXiv:2603.19020 [gr-qc]} \BibitemShut {NoStop}%
\bibitem [{\citenamefont {Ma}\ \emph {et~al.}(2024)\citenamefont {Ma},
  \citenamefont {Niu},\ and\ \citenamefont {Zhao}}]{Ma:2024kkz}%
  \BibitemOpen
  \bibfield  {author} {\bibinfo {author} {\bibfnamefont {Z.-C.}\ \bibnamefont
  {Ma}}, \bibinfo {author} {\bibfnamefont {R.}~\bibnamefont {Niu}},\ and\
  \bibinfo {author} {\bibfnamefont {W.}~\bibnamefont {Zhao}},\ }\bibfield
  {title} {\bibinfo {title} {{The Multi-parameter Test of Gravitational Wave
  Dispersion with Principal Component Analysis}},\ }\href
  {https://doi.org/10.1088/1674-4527/ad3c70} {\bibfield  {journal} {\bibinfo
  {journal} {Res. Astron. Astrophys.}\ }\textbf {\bibinfo {volume} {24}},\
  \bibinfo {pages} {055012} (\bibinfo {year} {2024})},\ \Eprint
  {https://arxiv.org/abs/2401.17666} {arXiv:2401.17666 [gr-qc]} \BibitemShut
  {NoStop}%
\bibitem [{\citenamefont {Shoom}\ \emph {et~al.}(2023)\citenamefont {Shoom},
  \citenamefont {Gupta}, \citenamefont {Krishnan}, \citenamefont {Nielsen},\
  and\ \citenamefont {Capano}}]{Shoom:2021mdj}%
  \BibitemOpen
  \bibfield  {author} {\bibinfo {author} {\bibfnamefont {A.~A.}\ \bibnamefont
  {Shoom}}, \bibinfo {author} {\bibfnamefont {P.~K.}\ \bibnamefont {Gupta}},
  \bibinfo {author} {\bibfnamefont {B.}~\bibnamefont {Krishnan}}, \bibinfo
  {author} {\bibfnamefont {A.~B.}\ \bibnamefont {Nielsen}},\ and\ \bibinfo
  {author} {\bibfnamefont {C.~D.}\ \bibnamefont {Capano}},\ }\bibfield  {title}
  {\bibinfo {title} {{Testing the post-Newtonian expansion with GW170817}},\
  }\href {https://doi.org/10.1007/s10714-023-03100-z} {\bibfield  {journal}
  {\bibinfo  {journal} {Gen. Rel. Grav.}\ }\textbf {\bibinfo {volume} {55}},\
  \bibinfo {pages} {55} (\bibinfo {year} {2023})},\ \Eprint
  {https://arxiv.org/abs/2105.02191} {arXiv:2105.02191 [gr-qc]} \BibitemShut
  {NoStop}%
\bibitem [{\citenamefont {Sathyaprakash}\ and\ \citenamefont
  {Dhurandhar}(1991)}]{Sathyaprakash:1991mt}%
  \BibitemOpen
  \bibfield  {author} {\bibinfo {author} {\bibfnamefont {B.~S.}\ \bibnamefont
  {Sathyaprakash}}\ and\ \bibinfo {author} {\bibfnamefont {S.~V.}\ \bibnamefont
  {Dhurandhar}},\ }\bibfield  {title} {\bibinfo {title} {{Choice of filters for
  the detection of gravitational waves from coalescing binaries}},\ }\href
  {https://doi.org/10.1103/PhysRevD.44.3819} {\bibfield  {journal} {\bibinfo
  {journal} {Phys. Rev. D}\ }\textbf {\bibinfo {volume} {44}},\ \bibinfo
  {pages} {3819} (\bibinfo {year} {1991})}\BibitemShut {NoStop}%
\bibitem [{\citenamefont {Damour}\ \emph {et~al.}(2000)\citenamefont {Damour},
  \citenamefont {Iyer},\ and\ \citenamefont {Sathyaprakash}}]{Damour:2000gg}%
  \BibitemOpen
  \bibfield  {author} {\bibinfo {author} {\bibfnamefont {T.}~\bibnamefont
  {Damour}}, \bibinfo {author} {\bibfnamefont {B.~R.}\ \bibnamefont {Iyer}},\
  and\ \bibinfo {author} {\bibfnamefont {B.~S.}\ \bibnamefont
  {Sathyaprakash}},\ }\bibfield  {title} {\bibinfo {title} {{Frequency domain P
  approximant filters for time truncated inspiral gravitational wave signals
  from compact binaries}},\ }\href {https://doi.org/10.1103/PhysRevD.62.084036}
  {\bibfield  {journal} {\bibinfo  {journal} {Phys. Rev. D}\ }\textbf {\bibinfo
  {volume} {62}},\ \bibinfo {pages} {084036} (\bibinfo {year} {2000})},\
  \Eprint {https://arxiv.org/abs/gr-qc/0001023} {arXiv:gr-qc/0001023}
  \BibitemShut {NoStop}%
\bibitem [{\citenamefont {Roy}\ \emph {et~al.}(2026)\citenamefont {Roy},
  \citenamefont {Haney}, \citenamefont {Pratten}, \citenamefont {T.~H.~Pang},\
  and\ \citenamefont {Van Den~Broeck}}]{Roy:2025gzv}%
  \BibitemOpen
  \bibfield  {author} {\bibinfo {author} {\bibfnamefont {S.}~\bibnamefont
  {Roy}}, \bibinfo {author} {\bibfnamefont {M.}~\bibnamefont {Haney}}, \bibinfo
  {author} {\bibfnamefont {G.}~\bibnamefont {Pratten}}, \bibinfo {author}
  {\bibfnamefont {P.}~\bibnamefont {T.~H.~Pang}},\ and\ \bibinfo {author}
  {\bibfnamefont {C.}~\bibnamefont {Van Den~Broeck}},\ }\bibfield  {title}
  {\bibinfo {title} {{Improved parametrized test of general relativity using
  the IMRPhenomX waveform family: Including higher harmonics and precession}},\
  }\href {https://doi.org/10.1103/855k-sys5} {\bibfield  {journal} {\bibinfo
  {journal} {Phys. Rev. D}\ }\textbf {\bibinfo {volume} {113}},\ \bibinfo
  {pages} {024016} (\bibinfo {year} {2026})},\ \Eprint
  {https://arxiv.org/abs/2504.21147} {arXiv:2504.21147 [gr-qc]} \BibitemShut
  {NoStop}%
\bibitem [{\citenamefont {S{\"a}nger}\ \emph {et~al.}(2026)\citenamefont
  {S{\"a}nger} \emph {et~al.}}]{Sanger:2024axs}%
  \BibitemOpen
  \bibfield  {author} {\bibinfo {author} {\bibfnamefont {E.~M.}\ \bibnamefont
  {S{\"a}nger}} \emph {et~al.},\ }\bibfield  {title} {\bibinfo {title} {{Tests
  of general relativity with GW230529: A neutron star merging with a lower
  mass-gap compact object}},\ }\href {https://doi.org/10.1103/r43k-51yq}
  {\bibfield  {journal} {\bibinfo  {journal} {Phys. Rev. D}\ }\textbf {\bibinfo
  {volume} {113}},\ \bibinfo {pages} {084070} (\bibinfo {year} {2026})},\
  \Eprint {https://arxiv.org/abs/2406.03568} {arXiv:2406.03568 [gr-qc]}
  \BibitemShut {NoStop}%
\bibitem [{\citenamefont {Green}\ and\ \citenamefont
  {Gair}(2021)}]{Green:2020dnx}%
  \BibitemOpen
  \bibfield  {author} {\bibinfo {author} {\bibfnamefont {S.~R.}\ \bibnamefont
  {Green}}\ and\ \bibinfo {author} {\bibfnamefont {J.}~\bibnamefont {Gair}},\
  }\bibfield  {title} {\bibinfo {title} {{Complete parameter inference for
  GW150914 using deep learning}},\ }\href
  {https://doi.org/10.1088/2632-2153/abfaed} {\bibfield  {journal} {\bibinfo
  {journal} {Mach. Learn. Sci. Tech.}\ }\textbf {\bibinfo {volume} {2}},\
  \bibinfo {pages} {03LT01} (\bibinfo {year} {2021})},\ \Eprint
  {https://arxiv.org/abs/2008.03312} {arXiv:2008.03312 [astro-ph.IM]}
  \BibitemShut {NoStop}%
\bibitem [{\citenamefont {Datta}\ \emph {et~al.}(2024)\citenamefont {Datta},
  \citenamefont {Saleem}, \citenamefont {Arun},\ and\ \citenamefont
  {Sathyaprakash}}]{Datta:2022izc}%
  \BibitemOpen
  \bibfield  {author} {\bibinfo {author} {\bibfnamefont {S.}~\bibnamefont
  {Datta}}, \bibinfo {author} {\bibfnamefont {M.}~\bibnamefont {Saleem}},
  \bibinfo {author} {\bibfnamefont {K.~G.}\ \bibnamefont {Arun}},\ and\
  \bibinfo {author} {\bibfnamefont {B.~S.}\ \bibnamefont {Sathyaprakash}},\
  }\bibfield  {title} {\bibinfo {title} {{Multiparameter tests of general
  relativity using a principle component analysis with next-generation
  gravitational-wave detectors}},\ }\href
  {https://doi.org/10.1103/PhysRevD.109.044036} {\bibfield  {journal} {\bibinfo
   {journal} {Phys. Rev. D}\ }\textbf {\bibinfo {volume} {109}},\ \bibinfo
  {pages} {044036} (\bibinfo {year} {2024})},\ \Eprint
  {https://arxiv.org/abs/2208.07757} {arXiv:2208.07757 [gr-qc]} \BibitemShut
  {NoStop}%
\bibitem [{\citenamefont {Kumar}\ \emph {et~al.}(2025)\citenamefont {Kumar},
  \citenamefont {Gupta},\ and\ \citenamefont {Sathyaprakash}}]{Kumar:2025jio}%
  \BibitemOpen
  \bibfield  {author} {\bibinfo {author} {\bibfnamefont {D.}~\bibnamefont
  {Kumar}}, \bibinfo {author} {\bibfnamefont {I.}~\bibnamefont {Gupta}},\ and\
  \bibinfo {author} {\bibfnamefont {B.}~\bibnamefont {Sathyaprakash}},\
  }\href@noop {} {\bibinfo {title} {{Accelerating parameter estimation for
  parameterized tests of general relativity with gravitational-wave
  observations}}} (\bibinfo {year} {2025}),\ \Eprint
  {https://arxiv.org/abs/2511.16879} {arXiv:2511.16879 [gr-qc]} \BibitemShut
  {NoStop}%
\bibitem [{\citenamefont {Abac}\ \emph
  {et~al.}(2026{\natexlab{b}})\citenamefont {Abac} \emph
  {et~al.}}]{LIGOScientific:2025wao}%
  \BibitemOpen
  \bibfield  {author} {\bibinfo {author} {\bibfnamefont {A.~G.}\ \bibnamefont
  {Abac}} \emph {et~al.} (\bibinfo {collaboration} {LIGO Scientific, Virgo,
  KAGRA}),\ }\bibfield  {title} {\bibinfo {title} {{Black Hole Spectroscopy and
  Tests of General Relativity with GW250114}},\ }\href
  {https://doi.org/10.1103/6c61-fm1n} {\bibfield  {journal} {\bibinfo
  {journal} {Phys. Rev. Lett.}\ }\textbf {\bibinfo {volume} {136}},\ \bibinfo
  {pages} {041403} (\bibinfo {year} {2026}{\natexlab{b}})},\ \Eprint
  {https://arxiv.org/abs/2509.08099} {arXiv:2509.08099 [gr-qc]} \BibitemShut
  {NoStop}%
\bibitem [{\citenamefont {Ashton}\ \emph {et~al.}(2025)\citenamefont {Ashton},
  \citenamefont {Talbot}, \citenamefont {Roy}, \citenamefont {Pratten},
  \citenamefont {Pang}, \citenamefont {Agathos}, \citenamefont {Baka},
  \citenamefont {Sänger}, \citenamefont {Mehta}, \citenamefont {Steinhoff},
  \citenamefont {Maggio}, \citenamefont {Ghosh}, \citenamefont {Vijaykumar},
  \citenamefont {Enficiaud},\ and\ \citenamefont
  {Pompili}}]{ashton_2025_15676285}%
  \BibitemOpen
  \bibfield  {author} {\bibinfo {author} {\bibfnamefont {G.}~\bibnamefont
  {Ashton}}, \bibinfo {author} {\bibfnamefont {C.}~\bibnamefont {Talbot}},
  \bibinfo {author} {\bibfnamefont {S.}~\bibnamefont {Roy}}, \bibinfo {author}
  {\bibfnamefont {G.}~\bibnamefont {Pratten}}, \bibinfo {author} {\bibfnamefont
  {T.-H.}\ \bibnamefont {Pang}}, \bibinfo {author} {\bibfnamefont
  {M.}~\bibnamefont {Agathos}}, \bibinfo {author} {\bibfnamefont
  {T.}~\bibnamefont {Baka}}, \bibinfo {author} {\bibfnamefont {E.}~\bibnamefont
  {Sänger}}, \bibinfo {author} {\bibfnamefont {A.}~\bibnamefont {Mehta}},
  \bibinfo {author} {\bibfnamefont {J.}~\bibnamefont {Steinhoff}}, \bibinfo
  {author} {\bibfnamefont {E.}~\bibnamefont {Maggio}}, \bibinfo {author}
  {\bibfnamefont {A.}~\bibnamefont {Ghosh}}, \bibinfo {author} {\bibfnamefont
  {A.}~\bibnamefont {Vijaykumar}}, \bibinfo {author} {\bibfnamefont
  {R.}~\bibnamefont {Enficiaud}},\ and\ \bibinfo {author} {\bibfnamefont
  {L.}~\bibnamefont {Pompili}},\ }\href
  {https://doi.org/10.5281/zenodo.15676285} {\bibinfo {title} {{Bilby TGR}}},\
  \bibinfo {howpublished} {Zenodo} (\bibinfo {year} {2025}),\ \bibinfo {note}
  {version v0.2}\BibitemShut {NoStop}%
\bibitem [{\citenamefont {{LIGO Scientific
  Collaboration}}(2016)}]{LIGOScientific:2016GW150914}%
  \BibitemOpen
  \bibfield  {author} {\bibinfo {author} {\bibnamefont {{LIGO Scientific
  Collaboration}}},\ }\href {https://doi.org/10.7935/K5MW2F23} {\bibinfo
  {title} {{Data release for event GW150914}}},\ \bibinfo {howpublished}
  {Gravitational Wave Open Science Center (GWOSC)} (\bibinfo {year}
  {2016})\BibitemShut {NoStop}%
\bibitem [{\citenamefont {Mezzasoma}\ \emph {et~al.}(2026)\citenamefont
  {Mezzasoma}, \citenamefont {Haster},\ and\ \citenamefont
  {Yunes}}]{Mezzasoma:2026wme}%
  \BibitemOpen
  \bibfield  {author} {\bibinfo {author} {\bibfnamefont {S.}~\bibnamefont
  {Mezzasoma}}, \bibinfo {author} {\bibfnamefont {C.-J.}\ \bibnamefont
  {Haster}},\ and\ \bibinfo {author} {\bibfnamefont {N.}~\bibnamefont
  {Yunes}},\ }\bibfield  {title} {\bibinfo {title} {{Impact of
  numerical-relativity waveform calibration on parametrized post-Einsteinian
  tests}},\ }\href {https://doi.org/10.1103/y6lc-fr77} {\bibfield  {journal}
  {\bibinfo  {journal} {Phys. Rev. D}\ }\textbf {\bibinfo {volume} {113}},\
  \bibinfo {pages} {124043} (\bibinfo {year} {2026})},\ \Eprint
  {https://arxiv.org/abs/2603.15765} {arXiv:2603.15765 [gr-qc]} \BibitemShut
  {NoStop}%
\bibitem [{\citenamefont {Kwok}\ \emph {et~al.}(2022)\citenamefont {Kwok},
  \citenamefont {Lo}, \citenamefont {Weinstein},\ and\ \citenamefont
  {Li}}]{Kwok:2021zny}%
  \BibitemOpen
  \bibfield  {author} {\bibinfo {author} {\bibfnamefont {J.~Y.~L.}\
  \bibnamefont {Kwok}}, \bibinfo {author} {\bibfnamefont {R.~K.~L.}\
  \bibnamefont {Lo}}, \bibinfo {author} {\bibfnamefont {A.~J.}\ \bibnamefont
  {Weinstein}},\ and\ \bibinfo {author} {\bibfnamefont {T.~G.~F.}\ \bibnamefont
  {Li}},\ }\bibfield  {title} {\bibinfo {title} {{Investigation of the effects
  of non-Gaussian noise transients and their mitigation in parameterized
  gravitational-wave tests of general relativity}},\ }\href
  {https://doi.org/10.1103/PhysRevD.105.024066} {\bibfield  {journal} {\bibinfo
   {journal} {Phys. Rev. D}\ }\textbf {\bibinfo {volume} {105}},\ \bibinfo
  {pages} {024066} (\bibinfo {year} {2022})},\ \Eprint
  {https://arxiv.org/abs/2109.07642} {arXiv:2109.07642 [gr-qc]} \BibitemShut
  {NoStop}%
\end{thebibliography}%

\end{document}